\documentclass[journal=ancac3,manuscript=article,layout=twocolumn]{achemso}
\usepackage{xcolor}
\usepackage[T1]{fontenc}
\usepackage[latin9]{inputenc}
\usepackage{amsmath}
\usepackage{amssymb}
\usepackage{bm}
\usepackage{babel}

\usepackage{hyperref}

\let\oldmaketitle\maketitle
\let\maketitle\relax

\title{A Pair of 2D Quantum Liquids: Investigating the Phase Behavior of Indirect Excitons}

\author{Paul R. Wrona}
\email{pwrona2@berkeley.edu}
\affiliation{Department of Chemistry, University of California, Berkeley, California 94720, USA}

\author{Eran Rabani}
\email{eran.rabani@berkeley.edu}
\affiliation{Department of Chemistry, University of California, Berkeley, California 94720, USA}
\alsoaffiliation{Materials Sciences Division, Lawrence Berkeley National Laboratory, Berkeley, California 94720, USA}
\alsoaffiliation{The Raymond and Beverly Sackler Center of Computational Molecular and Materials Science, Tel Aviv University, Tel Aviv 69978, Israel}

\author{Phillip L. Geissler\footnote{Phillip L. Geissler passed away on July 17, 2022, while this paper was under review.}{{~}$^{,}$}}
\affiliation{Department of Chemistry, University of California, Berkeley, California 94720, USA}
\alsoaffiliation{Materials Sciences Division, Lawrence Berkeley National Laboratory, Berkeley, California 94720, USA}

\begin{document}
\twocolumn[
\begin{@twocolumnfalse}
\oldmaketitle
\begin{abstract}
Long-lived indirect excitons (IXs) exhibit a rich phase diagram, including a Bose-Einstein condensate (BEC), a Wigner crystal, and other exotic phases. Recent experiments have hinted at a ``classical'' liquid of IXs above the BEC transition. To uncover the nature of this phase, we use a broad range of theoretical tools and find no evidence of a driving force towards classical condensation. Instead, we attribute the condensed phase to a quantum electron-hole liquid (EHL), first proposed by Keldysh for direct excitons. Taking into account the association of free carriers into bound excitons, we study the phase equilibrium between a gas of excitons, a gas of free carriers, and an EHL for a wide range of electron-hole separations, temperatures, densities, and mass ratios. Our results agree reasonably well with recent measurements of GaAs/AlGaAs coupled quantum wells.
\end{abstract}

\section{Keywords}
\textit{electron-hole liquid, indirect exciton, coupled quantum wells, phase transition, Mott transition}

\end{@twocolumnfalse}]

\section{Introduction}
Excitons are bound states of an electron and a hole attracted to each other by the screened electrostatic Coulomb force, resulting in neutral quasiparticles that can exist in a variety of semiconducting and insulating materials. Their lifetime is determined by the rate of decay to the ground state, either radiatively by emitting photons or non-radiatively by coupling to lattice phonons or other carriers via Auger recombination leading to exciton-exciton annihilation. Understanding these relaxation pathways has been key in the development of light-harvesting devices under low and high photon fluences.

\begin{figure}[!h]
\includegraphics[width=8cm]{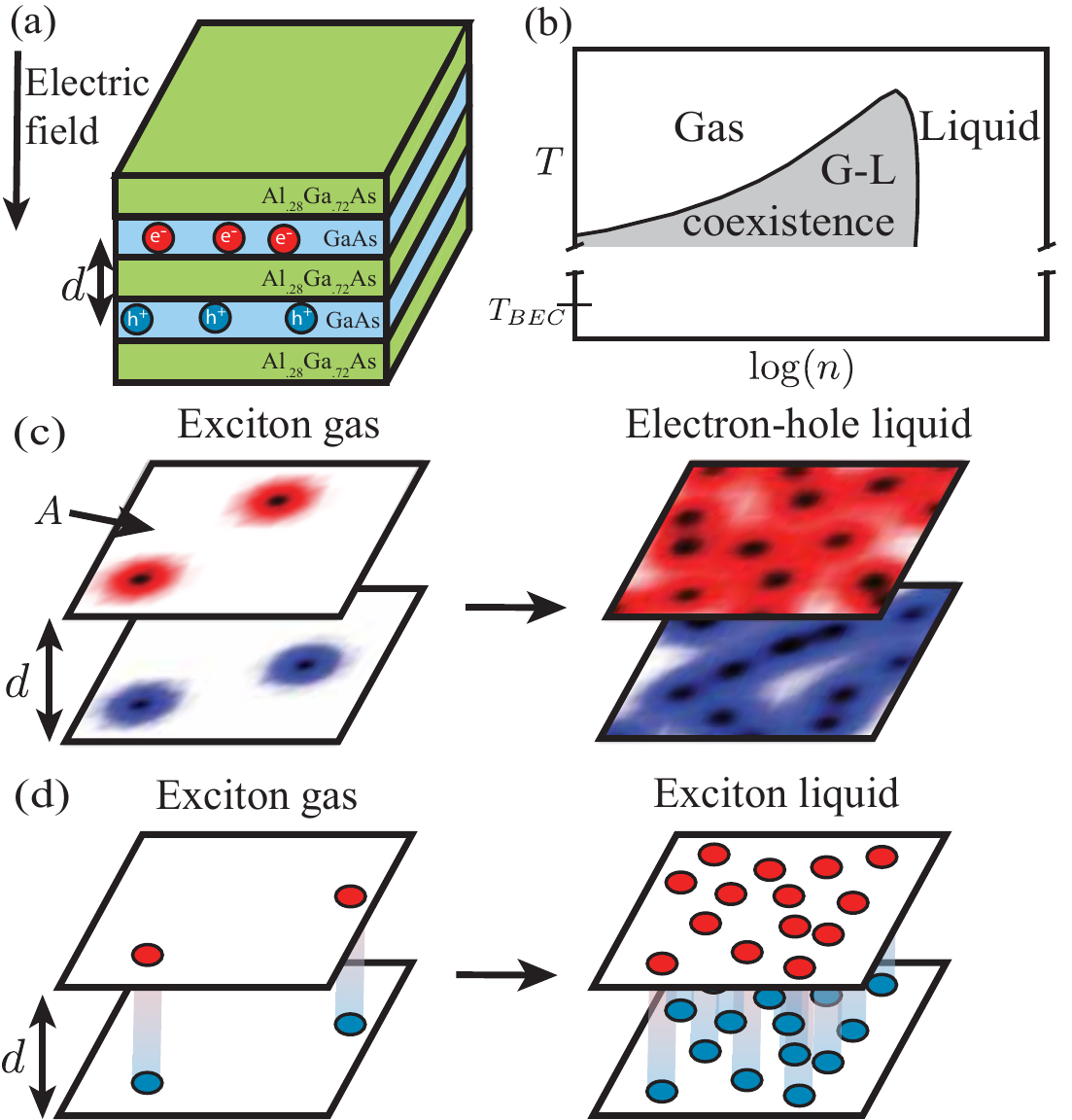}
\caption{\small (a) Schematic of coupled quantum wells with the center-to-center distance $d$ shown. (b) Sketch of a phase diagram showing the ordering of $T_{\rm BEC}$ and $T_{\rm C}$, the liquid-gas critical temperature for the transition observed by Bar-Joseph and co-workers. (c) Sketch of a phase transition from a gas of excitons to a degenerate electron-hole liquid. The area of each two-dimensional (2D) plane is $A$. (d) Sketch of a phase transition from a gas of bound indirect excitons to a classical liquid.}
\label{fig:sketch}
\end{figure}

The interactions among excitons can also result in a wide variety of thermodynamic phases. At high densities, excitons undergo a Mott transition to an electron-hole plasma stabilized by strong screening effects.\cite{key-1} At lower temperatures where quantum statistics dominate, the Mott transition is further facilitated by the favorable exchange interaction between like particles. Additionally, excitons may be regarded as weakly interacting neutral bosons~\cite{key-2} and can thus form Bose-Einstein condensates\textcolor{black}{\cite{key-3, key-32, key-31}} (BECs) and superfluid phases.\textcolor{black}{\cite{key-30}} However, the transience of excitons often complicates experimental realization of such quantum phases. After reaching thermal equilibrium, excitons eventually recombine (radiatively or non-radiatively), preventing further study of their phase behaviour. To prevent fast recombination of excitons, recent work has focused on indirect excitons (IXs), whose constituent carriers are confined to two parallel wells that are extended in two directions, due to either an electric field\cite{key-1} or type-II band alignment\cite{key-5} (see Fig.~\ref{fig:sketch}a). By restricting the carriers to different regions, IX recombination lifetimes are extended by orders of magnitude, providing a platform to better understand the phase behavior of excitons.


The interactions and collective behavior of indirect excitons differ significantly from their direct counterparts due to the permanent dipole moments they acquire through spatial separation of electrons and holes.
\cite{key-6} Theoretical studies of such dipolar fluids have revealed quantum and classical phases governed by intriguing correlation regimes.\cite{key-11,key-9,key-7,key-8,key-10} Indeed, several experimental studies have provided evidence for the formation of BECs of IXs at very low temperatures, typically below $1\mbox{K}$.\textcolor{black}{\cite{key-4,key-12}}  More recently, Bar-Joseph and his co-workers studied the collective behavior of IXs in GaAs/AlGaAs coupled quantum wells (CQWs) over a wider range of temperatures. 
Above the BEC temperature $T_{\rm BEC}\approx1.1\mbox{K}$ but below a critical temperature $T_{\rm C}=4.8\mbox{K}$, the excitons separated into two phases (see Fig.~\ref{fig:sketch}b) distinguished by a several-fold difference in exciton density, \textit{i.e.} a gas and a liquid, and characterized by a low-energy feature in the photoluminescence spectrum (the Z-line).
\cite{key-13} Based on previous theoretical work,\cite{key-7,key-8} they argued that the liquid phase 
results from
the repulsive interaction between the dipolar excitons,
which generate 
short-range correlations typical of a ``classical'' liquid. In a subsequent study,\cite{key-14} they concluded that the classical liquid is dark and the appearance of the Z-line in the photoluminescence spectrum is not due to recombination of excitons in the liquid phase, but rather, recombination of excitons in the gas phase near the interface with the liquid.\cite{key-14}
A dark exciton liquid was also observed by Rapaport and coworkers,\cite{key-15} consisting of electrons and holes in parallel spin configurations that cannot couple to light. Despite significant progress in our understanding of the phase behavior of IXs, the origin of the stability of the higher temperature classical liquid still remains unclear.

In this work, we revisit the putative classical liquid phase of excitons.  We seek to understand the nature of the liquid phase and the leading correlations that stabilize it at temperatures above $T_{\rm BEC}$ and below $T_{\rm C}$. In particular, we thoroughly assess whether a dense fluid of IXs, characterized by short-range correlations due to inter-exciton interactions, can coexist with a much more dilute gas of IXs at equilibrium.
The abrupt condensation implied by such a coexistence scenario is unlikely for spatially direct excitons, whose strong tendency to pair up generates at appreciable density a population of weakly interacting biexcitons, akin to a collection of diatomic hydrogen molecules that condense only at very low temperature. The transversely aligned dipoles of IXs, however, inhibit the formation of ``excitonic molecules'' \textcolor{black}{for large electron-hole separations where the exciton--exciton potential is purely repulsive. However, at moderate  separations, this potential is attractive,} raising the possibility that IXs condense through van der Waals-like interactions while remaining distinct -- neither paired as biexcitons nor strongly influenced by quantum degeneracy. The first part of our work examines this possibility by computing effective interaction potentials for pairs and triads of IXs, with approaches adapted from standard methods of quantum chemistry. Based on these calculations, we conclude that, for experimentally relevant values of the electron-hole separation, \textcolor{black}{excitons are repulsive species which lack an adhesive force that could drive classical condensation.}

The second part of our work explores an alternative interpretation of the liquid phase of IXs observed in experiments. It is well-known that a gas of spatially direct excitons can condense to form a degenerate electron-hole liquid (EHL), a plasma stabilized by spatial correlations in charge density, as first
proposed by Keldysh.\cite{key-16} Experiments in the following decades revealed interesting properties of this state, such as high mobility and simple mechanical control through applied stress.\cite{key-17,key-18,key-19,key-20} \textcolor{black}{This condensed state exists at temperatures low enough to achieve degeneracy, but not low enough to exhibit coherent phenomena or form a BEC.} Whether this EHL can account for the liquid phase of IXs is our second main focus.
For many different semiconductors, the critical temperature of Keldysh's EHL can be approximated by $T_{C}\approx0.1E_{\rm ex}/k_{\rm B}\label{eq:tc-ex formula}$, where $E_{\rm ex}$ is the binding energy of the exciton and $k_{\rm B}$ is Boltzmann's constant. To estimate $E_{\rm ex}$, one can model IXs with a bilayer geometry shown in Fig.~\ref{fig:sketch}c: electrons and holes are placed on two infinitely-thin parallel planes separated by $d$. 
We expect that the neglected out-of-plane fluctuations of the carriers are weak due to spatial confinement. For the experimental setup of Bar-Joseph,\cite{key-13,key-14} this model suggests a critical temperature of $T_{\rm C}=3.5\mbox{K}$ in comparison to the experimental value of $T_{\rm C}=4.8\mbox{K}$. This rough agreement suggests that this phase could be Keldysh's EHL realized in a bilayer geometry, as sketched in Fig.~\ref{fig:sketch}c. To study the Keldysh EHL phase and its dependence on the separation between electrons and holes, we adopt a Green's function approach and approximate the in-plane charge density fluctuations using the random phase approximation (RPA). 
We find that the Keldysh EHL is stable across a surprisingly wide range of planar separations, supporting the existence of a 
liquid of dissociated IXs that features strong screening and exchange interactions, rather than a classical liquid stabilized by cohesive forces between charge-neutral excitons.

\begin{figure}[!h]
\includegraphics[width=8cm]{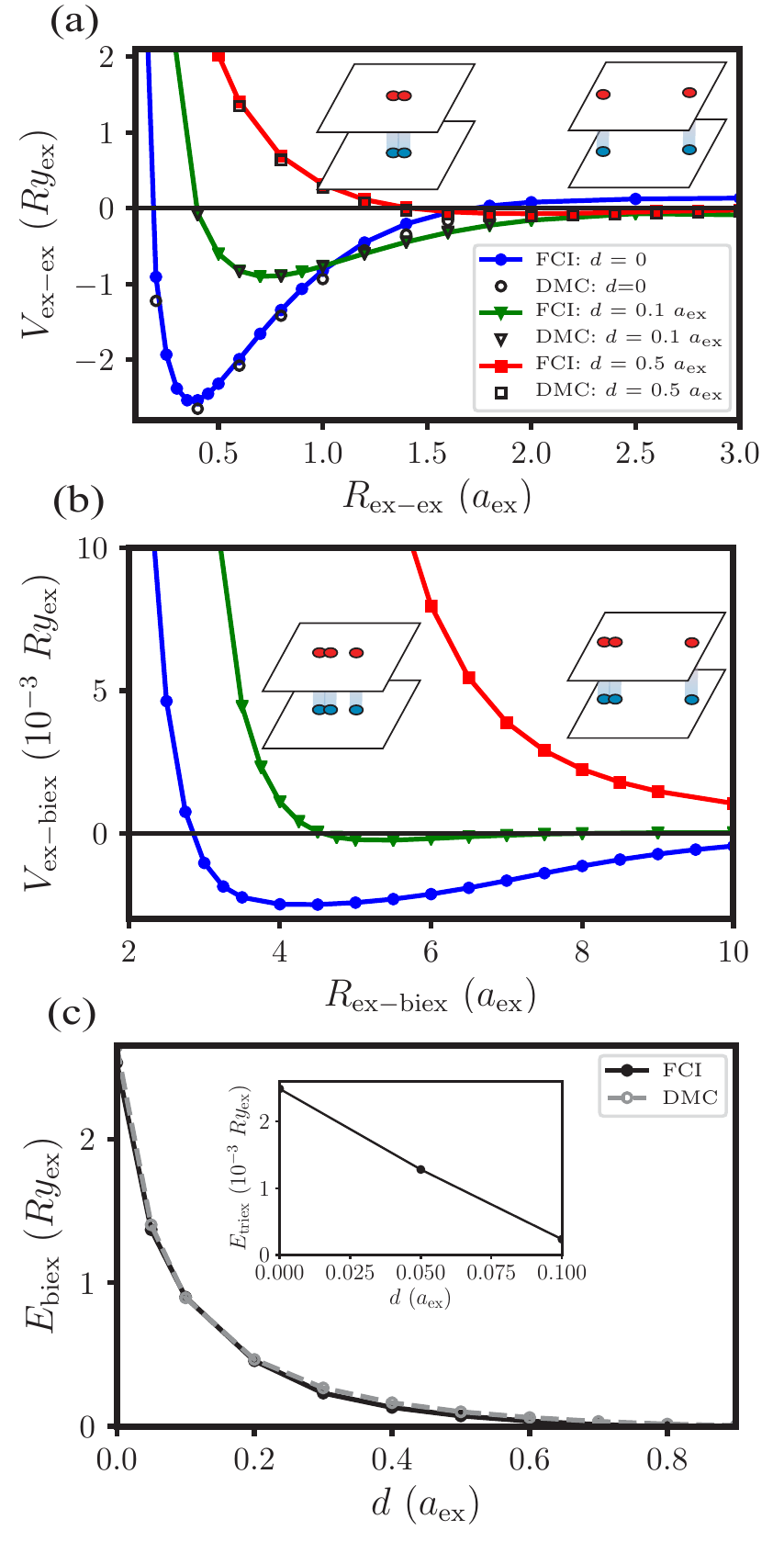}
\caption{\small (a) Interaction potentials between two excitons with infinitely heavy holes for various bilayer separations, $d$. We fix the orbitals' sizes as we pull apart the excitons, so we do not correctly describe dissociation. (See SI for further details.) ``FCI'' data come from our full CI method, and ``DMC'' data were computed using the CASINO program.\cite{key-22} (b) Interaction potentials between an exciton and a biexciton in a collinear geometry with infinitely heavy holes for various $d$, computed using our FCI method. $R_{\rm ex-biex}$ is the distance between the biexciton's center of mass and the third exciton. (c) Comparison of diffusion Monte Carlo results for the binding energy $E_{\rm biex}$ of a biexciton against this work's FCI method. The inset shows $E_{\rm triex}$, the binding energy of a triexciton.}
\label{fig:FCI}
\end{figure}

\section{Results and Discussion}
\subsection{Classical liquid}
Condensation of a classical fluid is typically driven by attractive interactions among its constituent particles. Purely repulsive interactions generate very high pressure at high particle densities; matching this pressure in a coexisting phase, as required for thermodynamic equilibrium, is difficult to achieve in a much more dilute state. A fluid of repulsive particles can of course undergo structural phase transitions, as famously exemplified by the crystallization of hard spheres. But coexisting phases of repulsive isometric particles are typically very similar in density, differing more prominently in symmetry or composition in the case of mixtures. Our scrutiny of the classical condensation hypothesis for IXs is thus principally a search for attractive interactions that could plausibly stabilize a dense phase of otherwise repulsive dipolar particles at moderate pressure.

We define an effective two-body interaction potential $V_{\rm ex-ex}(R_{\rm ex-ex})$ as the energy of two interacting excitons separated by a distance $R_{\rm ex-ex}$ minus the energy of two non-interacting excitons ($-2E_{\rm ex}$). This requires a Born-Oppenheimer-like approximation, in essence taking $R_{\rm ex-ex}$ to be fixed while averaging over quantum fluctuations in the excitons' internal structure.
Justifying this simplification requires that the hole is much more (or much less) massive than its partner electron. Our calculation of excitonic interaction potentials will therefore assume infinitely massive holes. In materials of interest, the electron-hole mass ratio $\sigma = m_{\rm e}/m_{\rm h}$ is not nearly so extreme. The heavy-hole limit ($\sigma=0$) we consider nonetheless provides a useful assessment, as it represents the most favorable scenario for attraction among excitons.

Interactions between a pair of IXs have been computed by Needs and co-workers\cite{key-21} using diffusion Monte Carlo (DMC) methods for the same bilayer Hamiltonian we consider. Their results reveal a two-body attraction that weakens rapidly with increasing separation $d$. To serve as a basis for classical condensation, this attraction would need to be additive, i.e., a similarly favorable energy would need to be realized as a third exciton is added, then a fourth, and so on. DMC is not well suited for evaluating this additivity, since the electron/hole wavefunction acquires nodal surfaces when $N>2$. We instead adopt a configuration interaction (CI) approach, improving systematically on a Hartree-Fock-like mean field approximation, just as in highly accurate quantum chemistry calculations. The SI describes our full CI method in detail, which assumes infinite hole mass and employs a suitable localized basis set. To demonstrate its accuracy, we show in Fig.~\ref{fig:FCI}a computed pair potentials $V_{\rm ex-ex}$ for several bilayer separations, together with DMC results computed using the CASINO program. \cite{key-22} While attraction between IXs remains evident at $d=0.5a_{\rm ex}$, the biexciton binding energy is a small fraction of $Ry_{\rm ex}$ at this separation. Near $d=0.8a_{\rm ex}$ the minimum of the interaction potential becomes too shallow to resolve, and for significantly larger $d$ the pair potential is purely repulsive.

For $d=0$, a hydrogenic analogy suggests that the exciton pair attraction represents a kind of covalent bond, with substantial sharing of electron density. As with diatomic hydrogen, we then expect that interactions between this biexciton and additional excitons are noncovalent in character and thus considerably weaker. The exciton-biexciton potential $V_{\rm ex-biex}$, plotted in Fig.~\ref{fig:FCI}b, verifies this expectation. A van der Waals-like attraction favors distances much larger than the ``covalent bond'' length, and the scale of attractive energy is smaller by three orders of magnitude. The same is true for $d=0.1a_{\rm ex}$, despite dipolar repulsion that might be imagined to inhibit exciton pairing. For $d\geq0.2a_{\rm ex}$ the effective potential $V_{\rm ex-biex}$ exhibits no minimum at all. The results for the biexction and triexciton binding energies as a function of the interlayer separation are summarized in Fig.~\ref{fig:FCI}c.

The attraction previously demonstrated between exciton pairs is thus not at all additive. Once paired, excitons experience at most extremely weak forces of cohesion. Experimentally relevant bilayer separations $d > 0.5a_{\rm ex}$ entirely negate attractions involving biexcitons, casting doubt on the classical condensation picture. There remains the possibility that the repulsion among excitons'/biexcitons' dipoles generates correlations which stabilize liquid-gas phase coexistence.\cite{key-7, key-23} 
We tested this notion by performing classical Monte Carlo simulations of particles in two dimensions that repel at long range with energy $\sim R^{-3}$ and additionally exclude volume at close range. (See SI for details.) 
Computed isotherms manifest freezing transitions at high density and pressure but otherwise show no sign of thermodynamic bistability that could be associated with fluid condensation. 

\subsection{Quantum liquid}

\begin{figure}[!h]
\includegraphics{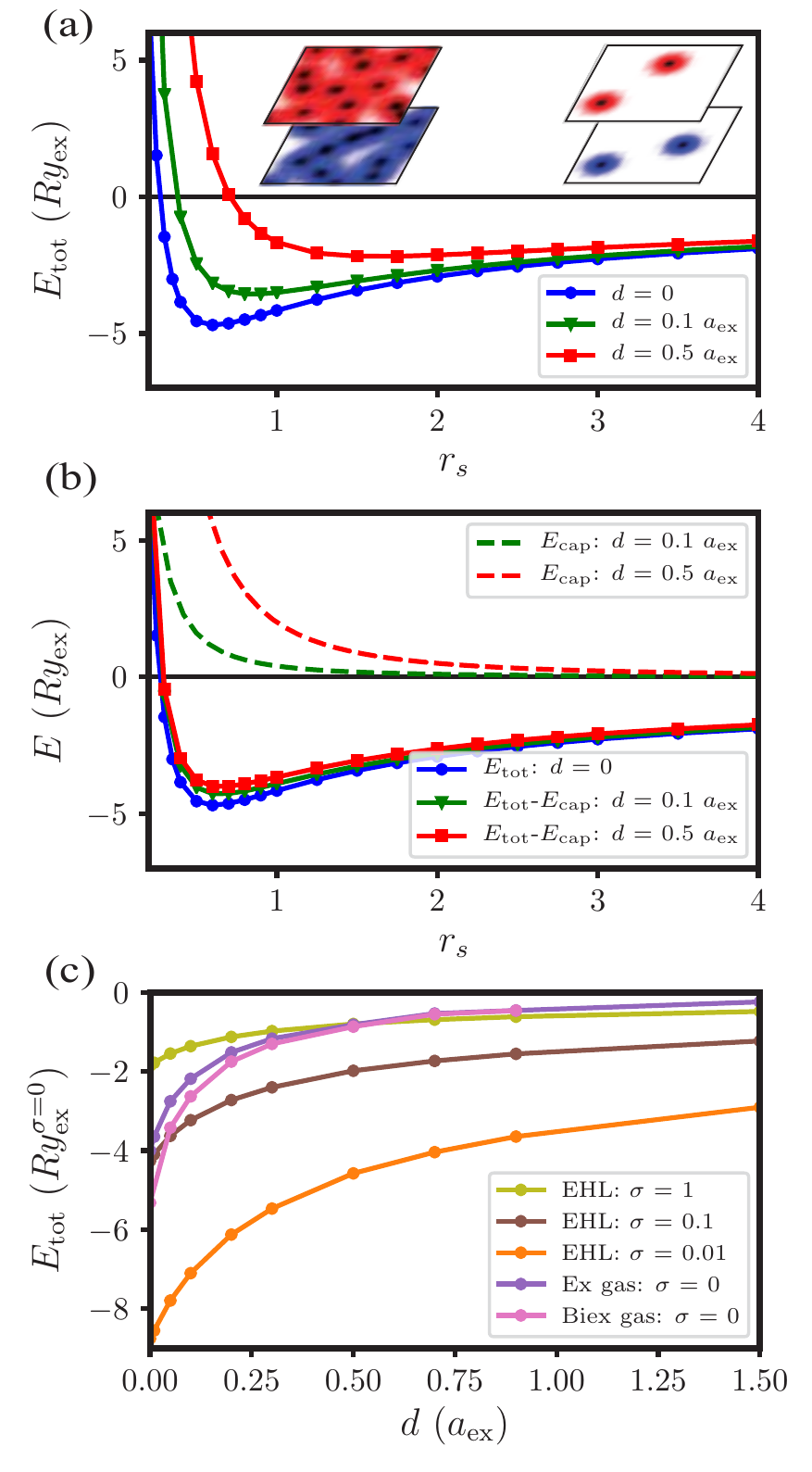}
\caption{\small (a) Total energy per electron of an EHL with $\sigma=0.1$ as a function of the
average interparticle spacing, $r_s$, evaluated for various $d$. (b) The
total energies shown in Fig.~\ref{fig:EHL energy}a minus the capacitor term (shown
in dashed lines) as a function of the average interparticle spacing.
(c) Minimum energy of an EHL with $\sigma=1$, 0.1, and 0.01. For the case $\sigma=0$, we also plot the energy of an exciton (``Ex'') gas at the same charge density as the EHL, and similarly for a biexciton (``Biex'') gas.} \label{fig:EHL energy}
\end{figure}

Turning to Keldysh's degenerate electron-hole liquid, we begin by considering the zero-temperature limit and focus on describing the relative stability of the EHL compared to the exciton gas. Finite temperature effects, including dissociation of bound excitons into an electron-hole gas, will be described below. The total energy per electron for $N$ electrons and $N$ holes is given by the sum of kinetic, exchange, capacitor, and correlation terms: $E_{\rm tot}=E_{\rm kin}+E_{\rm exch}+E_{\rm cap}+E_{\rm corr}$, where 
\begin{equation}
E_{\rm kin}=\frac{Ry_{\rm ex}}{r_{s}^{2}}
\end{equation}
and the dimensionless interparticle spacing, $r_{s}$, is determined by the relation\cite{key-24}
\begin{equation}
r_{s}^{2} =\frac{A}{\pi N a_{\rm ex}^2 },\label{eq: rs}
\end{equation}
where $A$ is the surface area depicted in Fig.~\ref{fig:sketch}. (See SI for all details.) In all of the calculations reported below, we take the thermodynamic limit, where $N\rightarrow\infty$ and $A\rightarrow\infty$, such that the number density, $n_{\rm tot}=N/A$, remains a constant. The exchange energy is given exactly by\cite{key-25}
\begin{equation}
E_{\rm exch}=-\frac{2.401}{r_{s}} Ry_{\rm ex}.
\end{equation}
The capacitor contribution (i.e., the classical electrostatic cost of separating uniformly charged plates by a perpendicular distance $d$), can be written as 
\begin{equation}
E_{\rm cap}=\frac{4d}{a_{\rm ex}} E_{\rm kin}. 
\end{equation}
Finally, the correlation energy in atomic units ($\hbar=1$) is estimated within the random phase approximation (RPA):
\begin{align}
E_{\rm corr}  =\frac{\sqrt{-1} A}{16 \pi^3} &\int d^{2}k\int d\omega \nonumber \\ & \times \int_{0}^{1}\frac{d\lambda}{\lambda}
{\bf \Pi}^T(k,\omega)~{\underline {\bf W}}~{\bf \Pi}(k,\omega),
\end{align}
where $\lambda$ is the coupling constant, ${\bf \Pi}^T(k,\omega)=\big[\Pi_{e}(k,\omega),\Pi_{h}(k,\omega)\big]$ and $\Pi_{i}(k,\omega)$ is the 2D Lindhard polarizability for particle $i={\rm e,h}$ evaluated at wavevector $k$ and frequency $\omega$. Within the RPA, the screened Coulomb matrix is given by:
\begin{align}
 {\underline {\bf W}} =
\left[\begin{array}{cc}
U_{\rm ee}U_{\rm ee}^{0} & U_{\rm eh}U_{\rm eh}^{0}\\ \\
U_{\rm eh}U_{\rm eh}^{0} & U_{\rm hh}U_{\rm hh}^{0}
\end{array}\right]
\end{align}
where $U_{ij}(\lambda,k,\omega)$ is the effective interaction between particles $i$ and $j$ and $U_{ij}^{0}(\lambda,k)$ is the corresponding bare Coulomb interaction in $k$-space. These quantities and the full details of RPA calculations are further described in the SI.

In Fig.~\ref{fig:EHL energy}a, we plot the resulting total energy per electron $E_{\rm tot}$ as a function of $r_{s}$ (Eq.~\ref{eq: rs}) for three different bilayer separations and for a mass ratio $\sigma=m_{e}/m_{h}=0.1$. We find that the total energy shows a pronounced minimum, $r_{\rm s,eq}$, near $a_{\rm ex}$ for small bilayer separations, signifying the existence of a stable degenerate electron-hole liquid. The major contribution to the change in the total energy as the bilayer separation increases is the capacitor term; without this term, results for different $d$ are nearly identical, as shown in Fig.~\ref{fig:EHL energy}b. \textcolor{black}{We note that the total energy of spatially separated electrons and holes has been calculated previously in the superfluid\cite{key-26} and superconducting\cite{key-27} regimes using a Green's function and variational approach, respectively. In both cases, the $d$-dependence on the energy agrees with our results.}

The minimum energies of an EHL with $\sigma=1$, $0.1$, and $0.01$ are shown in Fig.~\ref{fig:EHL energy}c. We also present the energies of a gas of excitons and gas of biexcitons. To compare these states on equal footing, we add to their energies the capacitor term evaluated at $r_{\rm s,eq}$. Ignoring the possibility of a BEC phase, we find that for most values of $d$, the EHL is the stable phase at zero-temperature and high densities. Specifically, for $\sigma=0.1$ and $d=1.5a_{\rm ex}$, we find that the total energy for carriers in the EHL is larger (i.e., more negative) than the energy of the IX gas by approximately 1 $Ry_{\rm ex}$. This value agrees with the observations of Bar-Joseph and co-workers,\cite{key-13} who measured a $\sim 1$ $Ry_{\rm ex}$ energy shift in their photoluminescence spectra between the IX gas and the condensed phase.

\begin{figure}[!h]
\includegraphics[width=8cm]{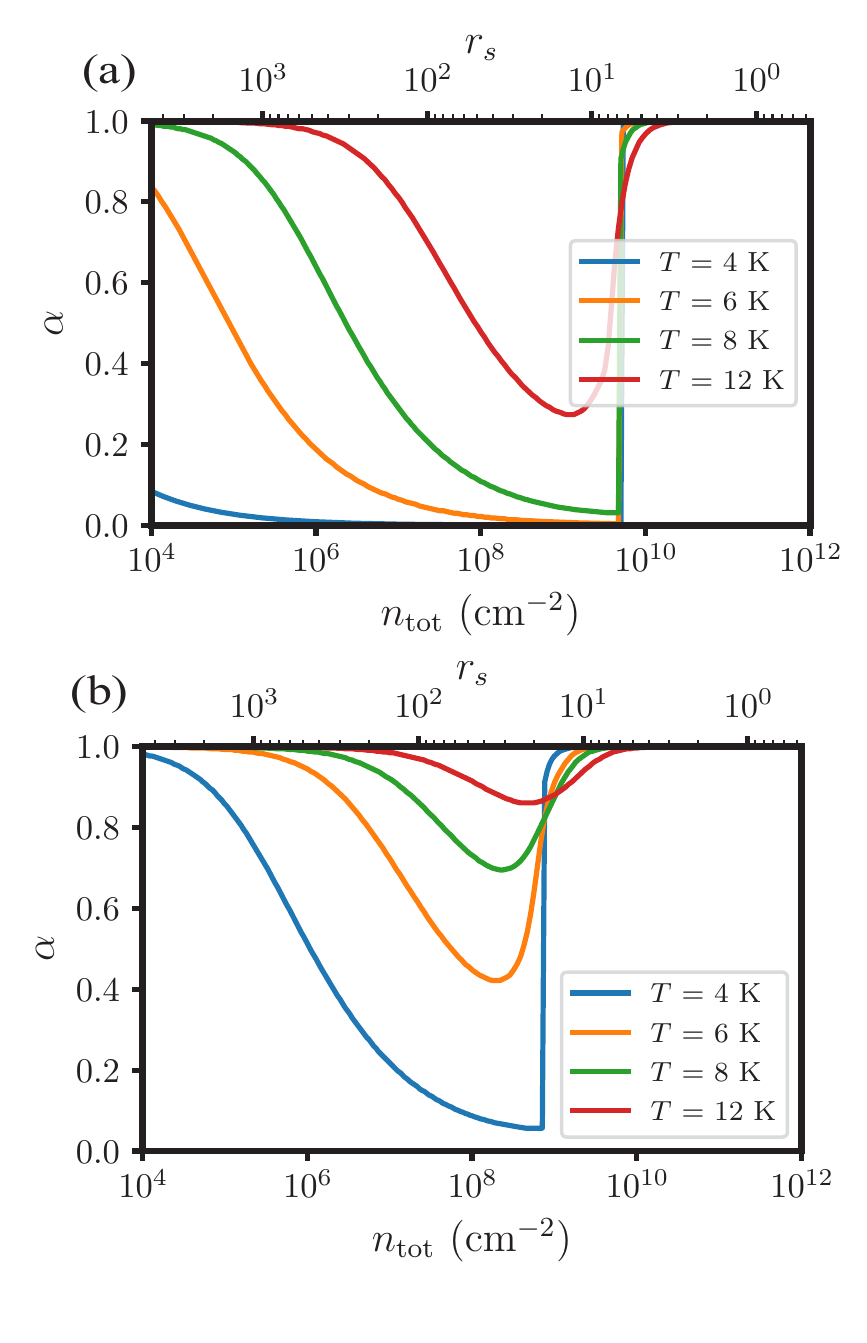}
\caption{\small Fraction of free carriers, $\alpha$, as a function of the total density $n_{\rm tot}$ (in units of excitations per cm$^{-2}$) for an electron-hole system with $\sigma=0.1$ at various temperatures. Results for $d=0.5$ $a_{\mathrm{ex}}$ are shown in (a), and for $d=1.5$ $a_{\mathrm{ex}}$ in (b).}
\label{fig:phase diagram}
\end{figure}

Next we turn to the effect of thermal fluctuations on the relative stability of IX gas and EHL phases. In doing so, it is important to acknowledge that the gas phase is not devoid of free charge carriers, nor is the liquid devoid of bound excitons. Instead, their proportions in each phase are determined by a chemical equilibrium ${\rm e}^{-}+{\rm h}^{+}\rightleftharpoons {\rm X}$ that requires the chemical potential $\mu_{\rm X}$ of an exciton to equal that of an unbound electron-hole pair, $\mu_{\rm eh}$. 
We treat interactions involving excitons as purely electrostatic and mean-field, giving
$\mu_{\rm X} = k_{\rm B}T \ln(1-\exp[-n_{\rm X}\lambda_{\rm X}^2/\xi_{\rm X}]) - E_{\rm ex} + \mu_{\rm cap}$, where the first term is an ideal contribution for bosons in two dimensions, $n_{\rm X}$ is the excitons' density, $\lambda_{\rm X}$ is their thermal de Broglie wavelength, $\xi_{\rm X}=4$ is their spin degeneracy, and $E_{\rm X}$ is their binding energy.
The capacitor potential, $\mu_{\rm cap}=4\pi de^2n_{\rm tot}$, depends only on $d$ and the total density $n_{\rm tot} = n_{\rm X} + n_{\rm eh}$ of excitations. 
The free carrier chemical potential,
\begin{align}
&\mu_{\rm eh} = k_{\rm B}T \ln[(\exp[n_{\rm eh}\lambda_{\rm e}^2/\xi_{\rm e}]-1) \nonumber \\  &\times (\exp[n_{\rm eh}\lambda_{\rm h}^2/\xi_{\rm h}])-1] + \mu_{\rm cap} + \mu_{\rm exch}^{\rm eh} + \mu_{\rm corr}^{\rm eh},
\label{eq:chempot}
\end{align}
includes an ideal contribution for the fermionic species ($\xi_{\rm e}=\xi_{\rm h}=2$), the capacitor potential, exchange effects from both electrons and holes, and a correlation term obtained from a generalization of RPA to finite temperature\cite{key-28} (see SI for details). We thus obtain a law of mass action for the fraction $\alpha=n_{\rm eh}/n_{\rm tot}$ of carriers that are not bound as excitons:
\eqref{eq:lawofmassaction}):
\begin{align}
K = &\frac{1-\exp[-n_{\rm tot}(1-\alpha) \lambda_{X}^{2} / \xi_{X}]} {\big(\exp[n_{\rm tot}\alpha\lambda_{\rm e}^{2}/\xi_{\rm e}]-1\big) \big(\exp[n_{\rm tot}\alpha\lambda_{\rm h}^{2}/\xi_{\rm h}]-1\big)},
\label{eq:lawofmassaction}
\end{align}
where
\begin{align}
K=\exp\big[\beta(E_{\rm ex}+\mu_{\rm exch}^{\rm eh}+\mu_{\rm corr}^{\rm eh})\big].
\label{eq:K}
\end{align}
For very low density ($n_{\rm tot} \ll \lambda_{\rm e}^{-2}$), effects of quantum statistics become unimportant, and Eq.~\ref{eq:lawofmassaction} reduces to the Saha ionization equation, a classical law of mass action. At densities typical of the degenerate EHL, quantum statistical effects are essential.

For a spatially uniform density $n_{\rm tot}$ of excitations, solving Eq.~\ref{eq:lawofmassaction} gives the fraction $\alpha$ of free carriers at thermal equilibrium. In Fig.~\ref{fig:phase diagram} we plot $\alpha$ as a function of $n_{\rm tot}$ for a) $d=0.5a_{\rm ex}$ and b) $d=1.5a_{\rm ex}$ for several temperatures, using parameters appropriate for GaAs. In the very dilute gas, $\alpha \approx 1 - (\text{constant})n e^{-\beta E_{\rm ex}}\approx 1$, since exciton dissociation is strongly favored by the entropy of mixing. With increasing density, $\alpha$ decreases steadily due to the favorable energy of exciton binding, until exchange and correlation become dominant at high density. $K$ rapidly approaches zero as a result, yielding a very small population of neutral excitons. Under some conditions the increase in free carrier fraction at high density occurs discontinuously, a result of Eq.~\ref{eq:lawofmassaction} acquiring multiple roots. (We select the root that minimizes the total free energy, as detailed in the SI.) This abrupt change in conductivity at density $n_{\rm Mott}(d,T)$  signals a first-order exciton Mott transition. It can occur only below a critical temperature $T_{\rm Mott}(d)$, as evident in Fig.~\ref{fig:phase diagram}(b) for $d=1.5a_{\rm ex}$ where $T_{\rm Mott}\approx 4$~K. As the exciton binding energy $E_{\rm ex}$ and EHL correlation energy $E_{\rm corr}$ decline in magnitude with increasing bilayer separation, $T_{\rm Mott}$ also decreases.

\begin{figure}[!h]
\includegraphics[width=8cm]{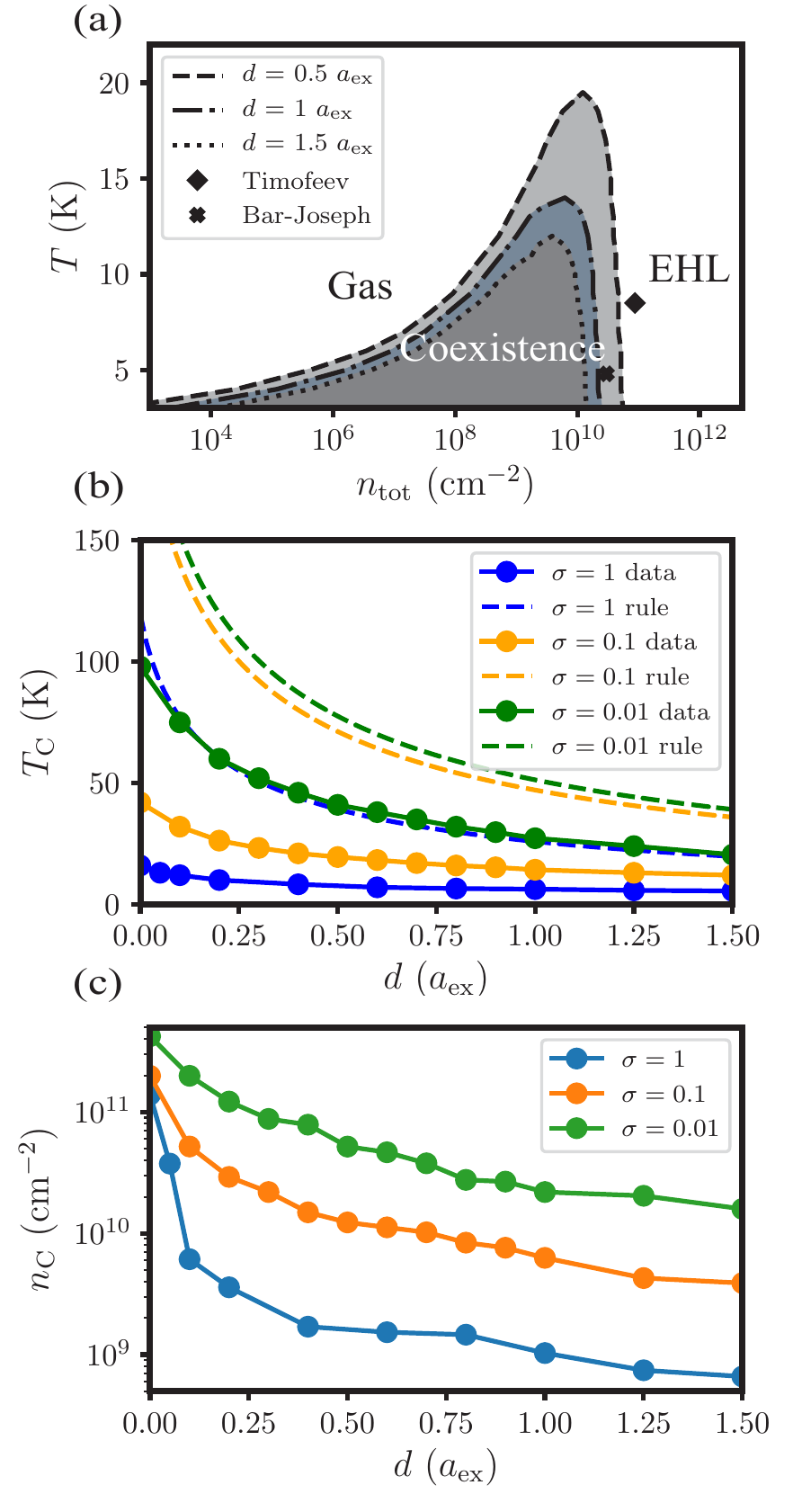}
\caption{\small (a) Phase diagrams in the density-temperature
plane for EHL condensation with $\sigma=0.1$ at various bilayer separations $d$. 
Experimentally estimated critical temperatures are shown for $d=1a_{\rm ex}$
(``Timofeev'' data \cite{key-29}) and for $d=1.5a_{\rm ex}$ (``Bar-Joseph'' data\cite{key-1, key-14}). The total density $n_{\rm tot}$ has units of excitations per cm$^{-2}$.
(b) Computed critical temperature as a function of bilayer separation $d$
for various mass ratios $\sigma.$ Also shown is the empirical rule $T_{C}\approx0.1E_{\rm ex}/k_{\rm B}$. (c) Computed critical (total) density as a function of bilayer separation $d$ for various mass ratios $\sigma.$}
\label{fig:critical points}
\end{figure}

The assumption of spatial uniformity, however, may break down before the Mott transition is encountered, and we find that this is in fact the case. The function $\mu_{\rm eh}(n_{\rm tot})$ we obtain by combining Eq.~\ref{eq:chempot} with the law of mass action develops an instability at low temperature. Specifically, $(\partial \mu_{\rm eh}/\partial n_{\rm tot})_T < 0$ over a range of intermediate densities, violating thermodynamic stability criteria and implying a phase-separated equilibrium state. We determine this state of coexistence -- typically featuring a low density gas enriched in IXs and a liquid of predominantly free carriers -- from $\mu_{\rm eh}(n_{\rm tot})$ using a standard Maxwell equal-area construction (see SI for details). Fig.~\ref{fig:critical points}a shows the resulting gas-EHL phase diagrams for three bilayer separations, each of which exhibits a first-order condensation transition below a critical temperature $T_{\rm C}(d)$. In each case, $T_{\rm C}$ exceeds $T_{\rm Mott}$, and the coexisting densities straddle $n_{\rm Mott}(d,T)$. States with uniform density $n_{\rm Mott}(d,T)$ are therefore unstable with respect to phase separation, and the first-order Mott transition described above is superseded by condensation.

With increasing bilayer separation, the critical excitation density $n_{\rm C}$ and temperature $T_{\rm C}$ both systematically decrease, as shown in Fig.~\ref{fig:critical points}b and c. The empirical formula $T_{\rm C}\approx0.1E_{\rm ex}/k_{B}$ anticipates this lowering of $T_{\rm C}$ due to the weakening exciton binding energy. We find that the phase diagrams for $d>0$ can be well approximated simply by adding the $d$-dependent capacitor term $\mu_{\rm cap}$ to the chemical potentials calculated for $d=0$ in addition to using the correct value of $E_{\rm ex}$.
Separating electrons and holes into distinct quantum wells thus appears to influence EHL condensation predominantly through a classical electrostatic bias, disfavoring dense-excitation states due to the necessity of separating substantial charge.

Fig.~\ref{fig:critical points}a also includes experimental data for GaAs, for which $\sigma = 0.1$ and $\epsilon = 12.9$. Beyond predicting the general decrease in $T_{\rm C}$ and $n_{\rm C}$ with increasing $d$, our estimates of the critical temperature are within a factor of 2-3 of the experimental data. While our predictions of $n_{\rm C}$ are off by an order of magnitude, we note that experimental measurements of the critical density often rely on mean-field or steady-state approximations and depend on many different parameters, such as the gate voltage. A material with different dielectric properties would set a different energy scale for electron-hole binding and screening, and the temperatures in Fig.~\ref{fig:critical points} would be scaled accordingly. A change in the mass ratio $\sigma$ has more subtle effects, but basic trends in $n_{\rm C}$ and $T_{\rm C}$ can be anticipated with the same reasoning used to explain EHL stability at zero-temperature. 
Because a very massive hole serves to localize electrons, we expect stabilization of the dense liquid phase with decreasing $\sigma < 1$ at fixed reduced mass (and similarly for increasing $\sigma > 1$). Correspondingly, $n_{\rm C}$ and $T_{\rm C}$ should both increase as $\sigma$ deviates from unity, as we observe and show in Fig. ~\ref{fig:critical points}b and c.

\section{Conclusions}
In summary, our approximate treatment of a simplified model for interacting electrons and holes in coupled quantum wells yields a low-temperature phase diagram that agrees reasonably well with experimental results. Given the assumptions we have made (a single band, effective masses, and a structureless isotropic background) and the experimental challenges of measuring a precise critical density and temperature, we consider the level of agreement to be a strong suggestion that the liquid phase observed in the laboratory has the same basic character as that in our model. Even for $d>1a_{\rm ex}$, the model's condensed phase is unambiguously a variant of Keldysh's electron-hole liquid -- a degenerate plasma of strongly screened charge carriers and very few bound excitons. By contrast, our calculation of effective interaction potentials among bound electron-hole pairs strongly discourages the notion of a classical liquid comprised of intact excitons as the equilibrium state at the temperatures and densities of interest. Cohesive forces that stabilize biexcitons weaken considerably as the bilayers separate, but they nonetheless dwarf any attraction to a third exciton. At $d = 1 a_{\rm ex}$ the interactions we compute are purely repulsive and cannot support phase coexistence between a sparse gas and dense liquid of excitons. The strong evidence for a stable Keldysh liquid of IXs and the predictions made for how the critical behavior changes with $d$ await experimental validation.

\section{Methods}
We analyze an idealized Hamiltonian $\hat{H}$ for a collection of $N$ electron-hole pairs in coupled quantum wells based on a single-band effective mass approximation. The total Hamiltonian in atomic units ($\hbar=m_0=e=4\pi\epsilon_0=1$, where $\hbar$ is the reduced Planck constant, $m_0$ is the electron's rest mass, $e$ is the elementary charge, and $\epsilon_0$ is the vacuum permittivity) reads:
\begin{equation}
\hat{H} = \hat{T} + \hat{V}
\label{eq: hamiltonian}
\end{equation}
where the kinetic energy is given by:
\begin{equation}
\textcolor{black}{\hat{T}=-\frac{1}{2m_{\rm e}}\sum_{i=1}^{N} \hat{\bf \nabla}_{{\rm e},i}^{2}-\frac{1}{2m_{\rm h}}\sum_{i=1}^{N} \hat{\bf \nabla}_{{\rm h},i}^{2}}
\label{eq:kinetic}
\end{equation}
and $m_{\rm e}$ $(m_{\rm h})$ is the electron (hole) effective mass.
Interactions among charge carriers are described by a screened Coulomb potential:
\begin{equation}
\begin{split}
\hat{V} = \frac{1}{\epsilon}\sum_{i=1}^{N}&\sum_{j>i}^{N}\frac{1}{|\hat{\bf r}_{{\rm e},i}-\hat{\bf r}_{{\rm e},j}|} + \frac{1}{\epsilon}\sum_{i=1}^{N}\sum_{j>i}^{N}\frac{1}{|\hat{\bf r}_{{\rm h},i}-\hat{\bf r}_{{\rm h},j}|} \\&-\frac{1}{\epsilon}\sum_{i=1}^{N}\sum_{j=1}^{N}\frac{1}{\sqrt{|\hat{\bf r}_{{\rm e},i}-\hat{\bf r}_{{\rm h},j}|^{2}+d^{2}}}.
\end{split}
\label{eq:potential}
\end{equation}
where $\epsilon$ is the static dielectric constant of the material. Excitonic units are used throughout this paper, with energies expressed relative to the exciton Rydberg $Ry_{\rm ex}=m_{\rm red}e^4/\big(2(4\pi\epsilon_0\epsilon)^2\hbar^2 \big)$ and lengths relative to the exciton Bohr radius $a_{\rm ex}=4\pi\epsilon_0\epsilon\hbar^2/(m_{\rm red}e^2)$, where $m_{\rm red}^{-1}=m_{\rm e}^{-1}+m_{\rm h}^{-1}$ is the electron-hole reduced mass.

\section{Associated Content}
This work was previously submitted to a pre-print server.\cite{key-33}
\subsection{Supporting Information}
The Supporting Information is available free of charge online. 

Derivation of single-particle basis sets; details of the FCI calculations; pressure-density isotherms for classical dipolar particles in two dimensions; derivation of contributions to the zero-temperature ground-state energy and to the Helmholtz free energy; fitting procedure for exchange-correlation free energies; procedure for solving the law of mass action; procedure for Maxwell equal-area constructions.

\section{Acknowledgments}
We would like to thank I. Bar-Joseph, R. Rapaport, and A. Rustagi for stimulating discussions.  E.R. is grateful to NSF-BSF International Collaboration in the Division of Materials Research program, NSF grant number DMR-2026741.

\end{document}


\section{Classical liquid}
\subsection{Single particle basis set}
Our single particle basis was the set of solutions to the time-independent Schr$\ddot{{\rm o}}$dinger equation for an exciton in a bilayer geometry (shown in Fig. 1c in the main text). Within the single-band isotropic effective mass approximation and in relative coordinates,
\begin{equation}
\Big(-\frac{1}{2m_{{\rm red}}}\nabla^{2}-\frac{1}{\epsilon\sqrt{r^{2}+d^{2}}}\Big)\phi(r,\theta)=E\phi(r,\theta).
\end{equation}
The electron-hole reduced mass is $m_{{\rm red}}^{-1}=m_{{\rm e}}^{-1}+m_{{\rm h}}^{-1}$, $\epsilon$ is the static dielectric constant, and $d$ is the bilayer separation. 
After separating variables, 
$\phi_{n,m}(r,\theta)=R_{n,m}(r)\Theta_{m}(\theta)$,
the angular component is
\begin{equation}
\Theta_{m}(\theta)=\frac{1}{\sqrt{2\pi}}\exp[im\theta]
\end{equation}
where $m$ is the angular quantum number and the radial component is the solution to
\begin{equation}
-\frac{1}{2m_{\rm red}r}\frac{d}{dr}\big(r\frac{dR_{n,m}}{dr}\big)+\big(\frac{m^{2}}{2m_{\rm red}r^{2}}-\frac{1}{\epsilon\sqrt{r^{2}+d^{2}}}\big)R_{n,m}(r)=E_{n,m}R_{n,m}(r)
\end{equation}
where $n$ is the principal quantum number. We solved the radial equation on a real-space grid using the Arnoldi method to obtain the first several radial wavefunctions $R_{n,m}(r)$. We approximated these functions by Gaussians or products of Gaussians and polynomials of $x$ and $y$ as is typical for Gaussian basis sets.\citep{SzaboOstlund}
For $m\neq0$, these functions are not peaked at the origin so we first remove a factor of $r^m$: 
\begin{equation}
f(r)=r^{-m}R(r).
\end{equation}
The resulting functions $f$ were then approximated by a linear combination of Gaussians by 
choosing coefficients $c_{i}$ and variances $\sigma_{i}^{2}$ to minimize
\begin{equation}
\int_{0}^{\infty}r\big[f(r)-\sum_{i=1}^{N_{g}}c_{i}g(r,\sigma_{i}^{2})\big]^{2}dr
\end{equation}
using the BFGS algorithm.\cite{BFGS} We found that five Gaussians ($N_{g}=5$) yields an error less than 1\% for all states up to and including $n=4$.

\subsection{Full configuration interaction calculations for biexcitons}
Defining an exciton-exciton interaction potential required us to invoke a Born-Oppenheimer-like approximation as discussed in the main text. Thus, we assumed the holes were infinitely massive point particles on one plane and sought the electronic states located on the other plane. We used the full configuration interaction (FCI) method and solved the generalized eigenvalue problem \begin{equation}
    HC=SC\varepsilon.
\end{equation}
$H$ is the Hamiltonian matrix corresponding to the biexciton Hamiltonian (Eqs.~(1)-(3) in the main text with $N=2$), $C$ is the matrix of coefficients, $S$ is the overlap matrix, and $\varepsilon$ is the vector of biexciton energies $E_{\rm XX}$.

Our FCI biexciton wavefunction was the product of an asymmetric singlet spin function and the following symmetric spatial component:
\begin{equation}
\Psi_{{\rm biex}}(\textbf{r}_{1},\textbf{r}_{2})=\sum_{i,j=1}^{N_{\rm orbitals}}c_{ij}\big(\phi_{i}(\textbf{r}_{1})\phi_{j}(\textbf{r}_{2})+\phi_{j}(\textbf{r}_{1})\phi_{i}(\textbf{r}_{2})\big).
\end{equation}
$\textbf{r}_1$ is the position of electron 1, $\phi_{i}$ is an orbital characterized by the quantum numbers $n$ and $m$ and centered directly above one of the holes, \textit{et cetera}. The exciton-exciton interaction potential was defined as
\begin{equation}
    V_{\rm ex-ex}(R_{\rm ex-ex})=E_{\rm XX}(R_{\rm ex-ex})-2E_{\rm ex}
\end{equation}
where $R_{\rm ex-ex}$ is the separation between the two holes and $E_{\rm ex}$ is the energy of an indirect exciton within our basis. When we used ``scaled'' orbitals (discussed below), the exciton energy was evaluated using these scaled orbitals.

\begin{figure}[!h]
\centering
\includegraphics[width=12cm]{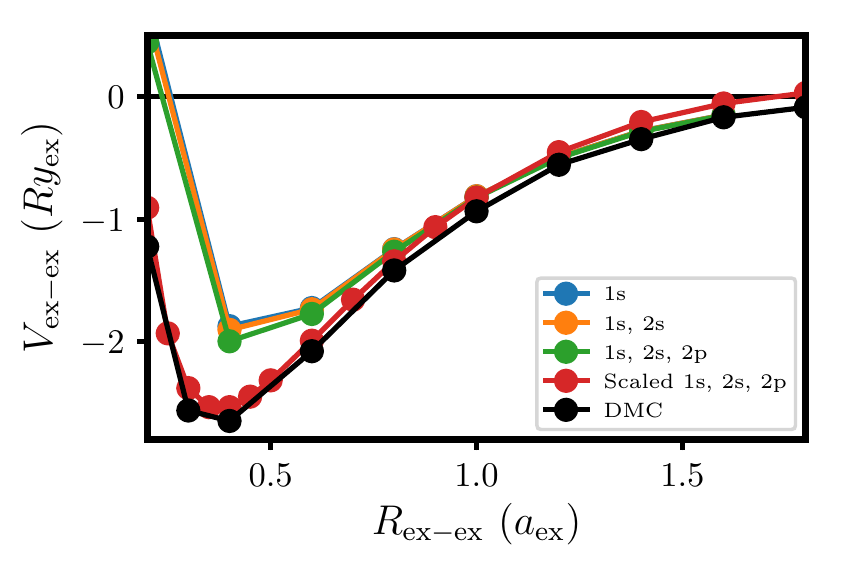}
\caption{Exciton-exciton interaction potentials for $d=0$. The legend contains the various single particle states used in the FCI calculations. The potentials converge towards the ``exact'' result (DMC) as the basis increases and improves in quality.}
\label{fig:s1}
\end{figure}

Fig.~\ref{fig:s1} shows exciton-exciton interaction potentials for $d=0$ using various single particle orbitals. As we increased our basis size $N_{\rm orbitals}$, both $E_{\rm ex}$ and
$E_{\rm XX}$ decreased. However, the latter energy almost always decreased more quickly, so $V_{\rm ex-ex}$ decreased as our basis set improved in general. Our basis set is complete in theory so we would recover the exact (\textit{i.e.}, DMC) result if we use an infinite number of orbitals. However, as seen in Fig.~\ref{fig:s1} the convergence rate is rather slow. 
Inspired by Dunning's correlation-consistent basis sets,\cite{Dunning} we optimized the size of the exciton orbitals in order to minimize $E_{\rm XX}$ more quickly. Specifically, we divided $\sigma_i$, the standard deviations of the Gaussians used to construct these orbitals, by ``scaling factors'' in order to minimize $E_{\rm XX}$ at the equilibrium exciton-exciton separation $R_{\rm eq}$ and subsequently fixed these scaled orbitals for all exciton-exciton separations. 
Because we did not allow the orbitals to relax as the excitons were separated from one another, we did not describe dissociation properly. While this introduced an artificial long-range repulsion, the most important feature of these interaction potentials was the inter-exciton attraction. Table \ref{opt coeff} contains the scaling factors used to optimize the exciton orbitals in addition to the values of $R_{\rm eq}$ for all bilayer separations under consideration.

\begin{table}
\centering{}%
\begin{tabular}{|c|c|c|c|c|c|}
\hline 
$d$ (a$_{\text{\rm ex}})$ & 1s factor & 2p$_{x}$ factor & 2p$_{y}$ factor & 2s factor  & $R_{\rm eq}$ ($a_{\rm ex}$)\tabularnewline
\hline 
\hline 
0 & 1.3 & 0.8  & 1.0 & 4.6 & 0.4\tabularnewline
\hline 
0.05 & 1.2 & 0.6 & 5.0 & 3.7 & 0.55\tabularnewline
\hline 
0.1  & 1.1 & 3.8 & 4.1 & 3.2 & 0.7\tabularnewline
\hline 
0.2 & 1.1 & 3.1 & 3.4 & 2.7 & 1.0\tabularnewline
\hline 
0.3 & 1.0 & 2.5 & 2.8 & 2.5 & 1.3\tabularnewline
\hline 
0.4 & 1.0 & 2.3 & 2.5 & 2.2 & 1.7\tabularnewline
\hline 
0.5 & 1.0 & 2.1 & 2.3 & 2.0 & 2.0\tabularnewline
\hline 
0.6 & 1.0 & 2.0 & 2.2 & 1.9 & 2.3\tabularnewline
\hline 
0.7 & 1.0 & 1.9 & 2.1 & 1.7 & 2.7\tabularnewline
\hline 
0.8 & 1.0 & 1.8 & 1.9 & 1.5 & 3.1\tabularnewline
\hline 
0.9 & 1.0 & 1.7 & 1.8 & 1.3 & 3.8\tabularnewline
\hline 
\end{tabular}
\caption{Optimal scaling factors for our Gaussian basis sets.}
\label{opt coeff}
\end{table}

\subsection{Full configuration interaction calculations for triexcitons}
We computed triexciton energies $E_{\rm XXX}$ in a manner similar to the biexciton case. The ground state spin configuration for three electrons is a doublet. To derive the corresponding spatial wavefunction, we started from the following Slater determinant:
\begin{equation}
\Psi_{XXX}(\textbf{r}_{1},\textbf{r}_{2},\textbf{r}_{3},\omega_{1},\omega_{2},\omega_{3})=\left|\begin{array}{ccc}
\phi_{a}(\textbf{r}_{1})\alpha(\omega_{1}) & \phi_{b}(\textbf{r}_{1})\beta(\omega_{1}) & \phi_{c}(\textbf{r}_{1})\alpha(\omega_{1})\\
\phi_{a}(\textbf{r}_{2})\alpha(\omega_{2}) & \phi_{b}(\textbf{r}_{2})\beta(\omega_{2}) & \phi_{c}(\textbf{r}_{2})\alpha(\omega_{2})\\
\phi_{a}(\textbf{r}_{3})\alpha(\omega_{3}) & \phi_{b}(\textbf{r}_{3})\beta(\omega_{3}) & \phi_{c}(\textbf{r}_{3})\alpha(\omega_{3})
\end{array}\right|
\label{triexciton Slater det}
\end{equation}
where $\phi_a$ is a single particle spatial orbital, $\alpha$ and $\beta$ are spin functions, and $\omega_{i}$ is the spin variable for electron $i$. Without loss of generality, we paired the spins of electrons 1 and 2 into a singlet state and coupled the spin of the third electron which we chose to be up. In the uncoupled basis, we have
\begin{equation}
\ket{\text{doublet}}=\Big(\frac{\ket{\uparrow\downarrow}_{1,2}-\ket{\downarrow\uparrow}_{1,2}}{\sqrt{2}}\Big)\ket{\uparrow}_{3}=\frac{\ket{\uparrow\downarrow\uparrow}_{1,2,3}-\ket{\downarrow\uparrow\uparrow}_{1,2,3}}{\sqrt{2}}
\label{triex uncoupled basis}
\end{equation}
Since singlet states have zero spin, we know that in the coupled basis,
\begin{equation}
    \ket{\text{doublet}}=\ket{\frac{1}{2};+\frac{1}{2}}
\end{equation}
Finally, we projected Eq.~(\ref{triexciton Slater det}) onto Eq.~(\ref{triex uncoupled basis}) to obtain
\begin{align}
 & \ket{\frac{1}{2};+\frac{1}{2}}\Big(\frac{\bra{\uparrow\downarrow\uparrow}-\bra{\downarrow\uparrow\uparrow}}{\sqrt{2}}\Big)\left|\begin{array}{ccc}
\phi_{a}(\textbf{r}_{1})\ket{\uparrow} & \phi_{b}(\textbf{r}_{1})\ket{\downarrow} & \phi_{c}(\textbf{r}_{1})\ket{\uparrow}\\
\phi_{a}(\textbf{r}_{2})\ket{\uparrow} & \phi_{b}(\textbf{r}_{2})\ket{\downarrow} & \phi_{c}(\textbf{r}_{2})\ket{\uparrow}\\
\phi_{a}(\textbf{r}_{3})\ket{\uparrow} & \phi_{b}(\textbf{r}_{3})\ket{\downarrow} & \phi_{c}(\textbf{r}_{3})\ket{\uparrow}
\end{array}\right| \nonumber \\
& =\Big(\phi_{a}(\textbf{r}_{1})\phi_{b}(\textbf{r}_{2})\phi_{c}(\textbf{r}_{3}) -\phi_{c}(\textbf{r}_{1})\phi_{b}(\textbf{r}_{2})\phi_{a}(\textbf{r}_{3}) \nonumber\\
&-\phi_{b}(\textbf{r}_{1})\phi_{c}(\textbf{r}_{2})\phi_{a}(\textbf{r}_{3})+\phi_{b}(\textbf{r}_{1})\phi_{a}(\textbf{r}_{2})\phi_{c}(\textbf{r}_{3})\Big) \frac{1}{\sqrt{2}}\ket{\frac{1}{2};+\frac{1}{2}} 
\end{align}
Summing over all possible orbitals, the FCI triexciton wavefunction reads
\begin{align}
    \Psi_{\rm triex}(\textbf{r}_{1},\textbf{r}_{2},\textbf{r}_{3}) & = \sum_{i,j}^{N_{\rm orbitals}} \sum_{k\neq i}^{N_{\rm orbitals}}\Big( \phi_{i}(\textbf{r}_{1})\phi_{j}(\textbf{r}_{2})\phi_{k}(\textbf{r}_{3})+\phi_{j}(\textbf{r}_{1})\phi_{i}(\textbf{r}_{2})\phi_{k}(\textbf{r}_{3}) \nonumber \\
&-\phi_{k}(\textbf{r}_{1})\phi_{j}(\textbf{r}_{2})\phi_{i}(\textbf{r}_{3})-\phi_{j}(\textbf{r}_{1})\phi_{k}(\textbf{r}_{2})\phi_{i}(\textbf{r}_{3})\Big)
\label{triex FCI wfn}
\end{align}

After computing triexciton energies, the exciton-biexciton interaction potential was computed according to
\begin{equation}
    V_{\rm ex-biex}(R_{\rm ex-biex};R_{\rm ex-ex})=E_{\rm XXX}(R_{\rm ex-biex})-E_{\rm ex}-E_{\rm XX}(R_{\rm ex-ex})
\end{equation}
$R_{\rm ex-ex}$ is the separation between the two closest excitons which we found to be almost exactly equal to $R_{\rm eq}$ in the energy-minimizing geometry. 
$R_{\rm ex-biex}$ is the separation between the biexciton's center-of-mass and the furthest exciton.

\subsection{Thermodynamics of dipolar particles in two dimensions}
Using computer simulations, we explored the possibility that quasiparticle correlations due to long-range dipolar repulsion can
drive condensation of a classical liquid in 2D. Specifically, we sampled particle configurations $r^N$ from a Boltzmann 
distribution $P(r^N)\propto e^{-\beta U(r^N)}$ with $U = (1/2)\sum_i \sum_{j\neq i} u(r_{ij})$. The pair interaction 
potential
$$
u(r) = 4\epsilon_{\rm LJ}\left[ 
\left({\sigma\over r}\right)^{12}
-\left({\sigma\over r}\right)^{6}
\right]
+ {A\over r^3},
$$
where $A$ is a positive constant, adds dipolar repulsion to the standard Lennard-Jones potential. Taking $\epsilon_{\rm LJ}$ and $\sigma$ as units of energy and length, respectively, the dimensionless potential becomes
$$
u(r) = 4\left(r^{-12} - r^{-6} \right)
+ {A^* \over r^3},
$$
where $A^* = A/(\epsilon_{\rm LJ}\sigma^3)$ sets the relative strength of dipolar repulsion.
For $A^* \gtrsim 2.2$ the classical potential $u(r)$ is repulsive at all $r$.
For $A^* \lesssim 2.2$, $u(r)$ has a local
minimum near contact, and for $A^* \lesssim 1.6$ this contact minimum becomes globally stable. 
The range $1.5 < A^* < 2.5$ thus spans the range of potentials we have computed using FCI methods.

To establish an appropriate range of temperature for this model system, 
we estimate the energy of dipolar repulsion near
contact for a coupled quantum well. In a bilayer geometry, a pair of excitons (each with the electron
perfectly aligned over the hole) separated by lateral distance $r$ has Coulomb interaction energy
$$
u_{\rm Coulomb}(r) = 2 {\rm Ry}_{\rm ex} \left[2 {a_{\rm ex}\over r}
- 2 {a_{\rm ex}\over \sqrt{r^2 + d^2}}  \right]
$$
which gives a dipolar repulsion to lowest order in $d/r$:
$$
u_{\rm dipole}(r) = 2 {\rm Ry}_{\rm ex} { (d/a_{\rm ex})^2\over
(r/a_{\rm ex})^3}
$$
Identifying $a_{\rm ex}$ as the rough volume-excluding size of an
exciton, and focusing on the case $d=a_{\rm ex}$ this gives an energy near contact of
$$
u_{\rm dipole}(r_{\rm contact}) = 2 {\rm Ry}_{\rm ex} .
$$
Comparing this energy scale to the observed critical temperature in GaAs,
$k_{\rm B} T_{\rm c} \approx 0.3 {\rm Ry}_{\rm ex}$, we have
$$
k_{\rm B} T_{\rm c} \approx 0.15 u_{\rm dipole}(r_{\rm contact})
$$
In our simple model, the dipolar energy at contact is simply $A/\sigma^3$, giving
$$
{k_{\rm B} T_{\rm c}\over \epsilon_{\rm LJ}} \approx 0.15 A^*.
$$
For the dipole strengths of interest ($1.5 < A^* < 2.5$), we would anticipate a
dimensionless critical temperature in the range $T_{\rm c} \approx 0.2$ to 0.4
(if condensation were to occur at all). 

Using Metropolis Monte Carlo sampling, we have computed pressure-density isotherms
at thermodynamic conditions spanning the ranges described above, for a 2D system
of $N=400$ particles. Specifically, 
we consider three dipole strengths, $A^* = 1.5, 2$, and 2.5. For each of these
strengths we consider three temperatures, $T = 0.1, 0.3$, and 0.5. Isotherms
are determined for each combination of $A^*$ and $T$
by calculating the average virial pressure at a series of densities,
beginning with a dilute gas at reduced density $\rho = 0.05$. The equilibrated state
at this density is then compressed to a slightly higher density $\rho' = \rho + \Delta\rho$,
with $\Delta\rho = 0.01$. Following an equilibration period of 1000 Monte Carlo sweeps, 
we compute a new average pressure over the course of 9000 sweeps. The density is increased
again by $\Delta\rho$, and equilibration and averaging are repeated as before. This 
compression protocol
is followed until the density reaches $\rho = 0.85$, a tightly packed and highly organized 
state with the rough appearance of crystalline order (which cannot strictly persist over
large scales in two dimensions). We subsequently reduce the density in steps of $\Delta\rho$,
computing a new series of average pressures at the same thermodynamic states as in the
compression protocol.

\begin{figure}[!h]
\centering
\includegraphics[width=7.5cm]{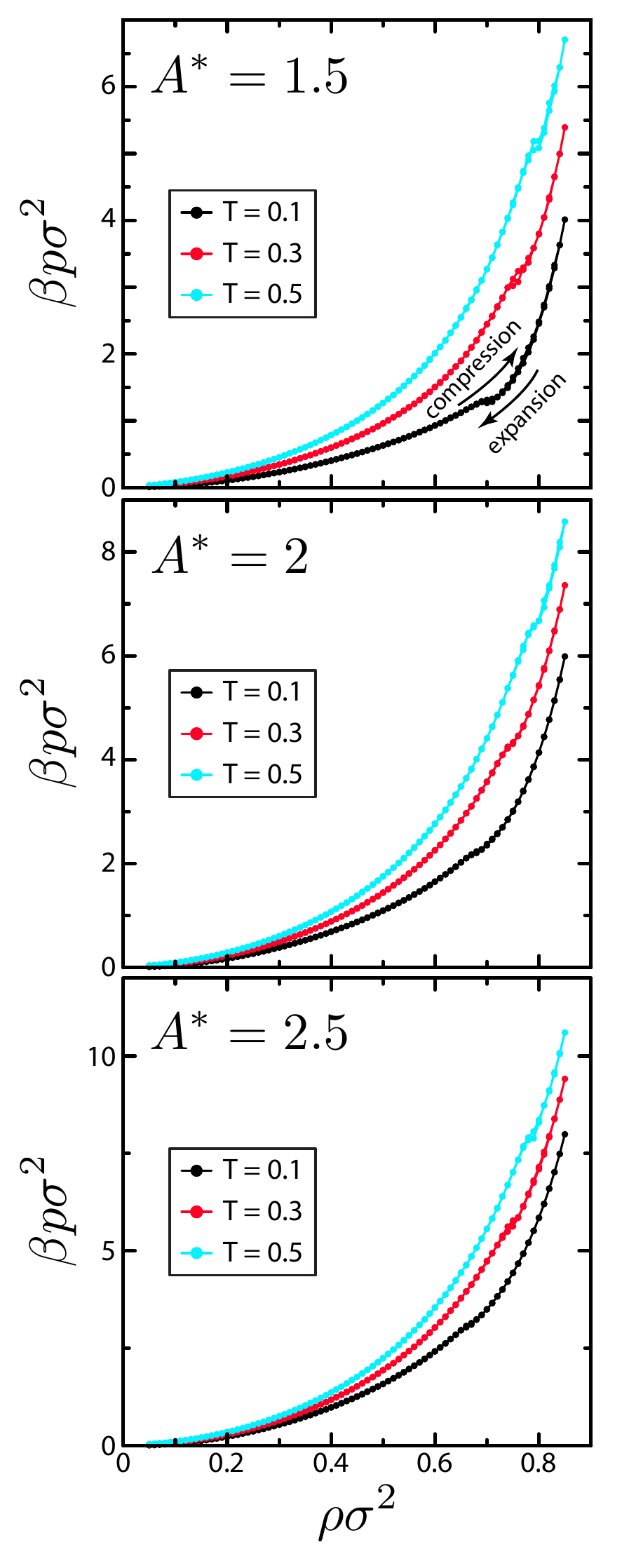}
\caption{Pressure-density isotherms for a system of dipolar particles
in two dimensions. For each dipole strength $A^*$, simulation results are shown
for three different temperatures. Each curve includes data for both compression and
expansion protocols. See text for additional numerical details.}
\label{fig:dipole-thermo}
\end{figure}

Computed isotherms are plotted in Fig.~\ref{fig:dipole-thermo}. A condensation transition would manifest in 
these results in two ways. First, it would
be accompanied by strong hysteresis, i.e., significant discrepancies between compression
and expansion results over a range of densities within the liquid-gas coexistence region,
due to high free energy barriers for condensation and vaporization. Second, the fully 
equilibrated macroscopic isotherm would feature constant pressure across the broad range of 
densities between gas and liquid. For our finite system, such nonanalytic behavior would be rounded
but still marked by a nearly constant pressure at coexistence. 
Neither of these features are evident in our results. 

Compression and expansion results are essentially indistinguishable at all but a few
densities. These small regions of hysteresis occur at high densities characteristic
of a solid phase. Given the subtleties of freezing in two dimensions,\cite{Bernard} we have not attempted 
to assign the high-density phase as hexatic or crystalline; for our purposes, it is
important only that (a) this state is not fluid, and (b) the fluid it coexists with is 
similarly dense, i.e., not a gas by any reasonable measure. The slight hysteresis
we find at high density is thus clearly associated with freezing rather than condensation.

In summary, the signatures of condensation -- hysteresis and near-constant pressure between
a low gas-like density and a much higher liquid-like density -- are entirely absent in
our simulation results. We note that a tail correction to the pressure,\cite{FrenkelSmit} resulting from
the unavoidable truncation of particle interactions, has not been included. This contribution,
proportional to the square of density, cannot give rise to hysteresis and would only 
exacerbate the observed non-constancy of 
pressure.


\section{Quantum liquid}
\subsection{Ground state energy at zero temperature}
Following previous work on the electron-hole liquid,\cite{BrinkmanRice, KuramotoKamimura} the ground state energy per electron at zero temperature was given by
\begin{equation}
    E_{\rm tot}=E_{\rm kin}+E_{\rm exch}+E_{\rm cap}+E_{\rm corr}.
\end{equation}

\subsubsection{Kinetic energy}
The total kinetic energy of all electrons and holes was expressed as the sum of the non-interacting energies over all occupied states:
\begin{equation}
\mathcal{E}_{\text{kin}}=\sum_{i=e,h}2\sum_{k<k_{\text{F}_{i}}}\frac{\hbar^2 k^{2}}{2m_{i}}
\end{equation}
where $k_{\text{F}_i}$ is the Fermi wavevector for particle $i$.
In the thermodynamic limit, the sum can be replaced by an integral, \cite{FetterWalecka} so
\begin{align}
\mathcal{E}_{\text{kin}} & =\sum_{i=\text{e,h}}2\frac{\hbar^2}{2m_{i}}\frac{A}{(2\pi)^{2}}\int_{0}^{k_{\text{F}_{i}}}2\pi k(k^{2})dk\\
 & =\sum_{i=\text{e,h}}2\frac{\hbar^2}{2m_{i}}\frac{A}{2\pi}\frac{k_{\text{F}_{i}}^{4}}{4}
\end{align}
After writing the Fermi wavevectors in terms of the density,
\begin{equation}
\mathcal{E}_{\text{kin}}=\sum_{i=\text{e,h}}2\frac{\hbar^2k_{\text{F}_{i}}^{2}}{2m_{i}}\frac{N}{4}
\end{equation}
\begin{equation}
\frac{\mathcal{E}_{\rm kin}}{N}=E_{\rm kin}=\hbar^2\Big(\frac{k_{\text{F}_{e}}^{2}}{2m_{e}}+\frac{ k_{\text{F}_{\rm h}}^{2}}{2m_{\rm h}}\Big)
\end{equation}
Using
\begin{equation}
    k_{\text{F}_i}a_{\rm ex}=\frac{\sqrt{2}}{r_{s}} \label{eq: kfaex}
\end{equation}
we have
\begin{equation}
    E_{\rm kin}=\frac{Ry_{\rm ex}}{r_s^2}
    \label{kin en}
\end{equation}

\subsubsection{Exchange energy}
The exchange energy was expressed as
\begin{equation}
    \mathcal{E}_{\text{\rm exch}}=-\sum_{i=\text{e,h}}\frac{2}{2A}\sum_{\textbf{k},\textbf{p}}U^0(\textbf{k}-\textbf{p})\Theta(k-k_{\text{F}_i})\Theta(p-k_{\text{F}_i})
\end{equation}
where the Fourier transform of the statically-screened two-dimensional Coulomb interaction is
\begin{equation}
    U^0(k)=\frac{2\pi e^2}{4\pi\epsilon_0\epsilon k}.
    \label{bare Coulomb}
\end{equation}
After replacing the double sum by a double integral, the expression was evaluated exactly: 
\begin{align}
\frac{\mathcal{E}_{\text{exch}}}{N} & = -\sum_{i=e,h}\frac{8\sqrt{2}}{3\pi r_s} Ry_{\rm ex}\\
 E_{\rm exch} & =-\frac{2.401}{r_{s}}Ry_{\rm ex} \label{exch en}
\end{align}

\subsubsection{Capacitor energy}
The capacitor energy is the $q=0$ term of the Fourier transform of the potential energy given in Eq.~3 in the main text. Each individual term in Eq.~3 diverges in the thermodynamic limit yet their sum is finite. We therefore introduced exponential convergence factors to regularize the terms:
\begin{equation}
V=\frac{1}{2}\sum_{i\neq j}^{N}\frac{\exp[-\mu|\textbf{r}_{e,i}-\textbf{r}_{e,j}|]}{\epsilon|\textbf{r}_{e,i}-\textbf{r}_{e,j}|}
+ \frac{1}{2}\sum_{i\neq j}^{N}\frac{\exp[-\mu|\textbf{r}_{h,i}-\textbf{r}_{h,j}|]}{\epsilon|\textbf{r}_{h,i}-\textbf{r}_{h,j}|} - \sum_{i,j}^{N} \frac{\exp[-\mu|\textbf{r}_{e,i}-\textbf{r}_{h,j}|]}{\epsilon \sqrt{ |\textbf{r}_{e,i}-\textbf{r}_{h,j}|^2 + d^2}}.
\end{equation}
To write this in second quantization, we first evaluated
\begin{align}
\bra{\textbf{k}_{1}\xi_{1}\textbf{k}_{2}\xi_{2}}V\ket{\textbf{k}_{3}\xi_{3}\textbf{k}_{4}\xi_{4}} & =\frac{1}{\epsilon A^{2}}\delta_{\xi_{1},\xi_{3}}\delta_{\xi_{2},\xi_{4}}\int d\textbf{r}_{1}\int d\textbf{r}_{2}\exp[-i(\textbf{k}_{1}\cdot \textbf{r}_{1}+\textbf{k}_{2}\cdot \textbf{r}_{2})]\\
 & \times\exp[i(\textbf{k}_{3}\cdot \textbf{r}_{1}+\textbf{k}_{4}\cdot \textbf{r}_{2})]\Big\{\frac{\exp[-\mu|\textbf{r}_{1}-\textbf{r}_{2}|]}{|\textbf{r}_{1}-\textbf{r}_{2}|}-\frac{\exp[-\mu|\textbf{r}_{1}-\textbf{r}_{2}|]}{\sqrt{|\textbf{r}_{1}-\textbf{r}_{2}|^{2}+d^{2}}}\Big\}\nonumber 
\end{align}
where $\xi_i$ is the spin variable of particle $i$.
Let $\textbf{r}=\textbf{r}_{2}$ and $\textbf{y}=\textbf{r}_{1}-\textbf{r}_{2}$, so
\begin{align}
\bra{\textbf{k}_{1}\xi_{1}\textbf{k}_{2}\xi_{2}}V\ket{\textbf{k}_{3}\xi_{3}\textbf{k}_{4}\xi_{4}} & =\frac{1}{\epsilon A^{2}}\delta_{\xi_{1},\xi_{3}}\delta_{\xi_{2},\xi_{4}}\int d\textbf{r}\exp[-i(\textbf{k}_{1}+\textbf{k}_{2}-\textbf{k}_{3}-\textbf{k}_{4})\cdot \textbf{r}]\\
 & \times\int d\textbf{y}\exp[i(\textbf{k}_{3}-\textbf{k}_{1})\cdot \textbf{y}]\Big\{\frac{\exp[-\mu y]}{y}-\frac{\exp[-\mu y]}{\sqrt{y^{2}+d^{2}}}\Big\}\nonumber
\end{align}
The $\textbf{r}$-integral gave an area times a Kronecker $\delta$-function which guarantees
momentum conservation. Let $\textbf{q}=\textbf{k}_{1}-\textbf{k}_{3}$ be the momentum transferred in the two-particle interaction.
\begin{align}
\bra{\textbf{k}_{1}\xi_{1}\textbf{k}_{2}\xi_{2}}V\ket{\textbf{k}_{3}\xi_{3}\textbf{k}_{4}\xi_{4}} & =\frac{1}{\epsilon A}\delta_{\xi_{1},\xi_{3}}\delta_{\xi_{2},\xi_{4}}\delta_{\textbf{k}_{1}+\textbf{k}_{2},\textbf{k}_{3}+\textbf{k}_{4}} \label{coulomb matrixele theta y} \\
& \times \int_{0}^{\infty}dy\int_{0}^{2\pi}d\theta y\exp[-iqy\cos\theta]\Big\{\frac{\exp[-\mu y]}{y}-\frac{\exp[-\mu y]}{\sqrt{y^{2}+d^{2}}}\Big\} \nonumber
\end{align}

To obtain the capacitor energy, we first address the $q=0$ case. After evaluating the $\theta$-integral,
\begin{align}
\bra{\textbf{k}_{1}\xi_{1}\textbf{k}_{2}\xi_{2}}V\ket{\textbf{k}_{3}\xi_{3}\textbf{k}_{4}\xi_{4}} & =\frac{1}{\epsilon A^{2}}\delta_{\xi_{1},\xi_{3}}\delta_{\xi_{2},\xi_{4}}\delta_{\textbf{k}_{1}+\textbf{k}_{2},\textbf{k}_{3}+\textbf{k}_{4}} \nonumber\\
    & \times\int_{0}^{\infty}dy\Big\{\exp[-\mu y]-\frac{y\exp[-\mu y]}{\sqrt{y^{2}+d^{2}}}\Big\}\quad \text{for }q=0
\label{q=0 term}
\end{align}
Turning to the $y$-integral, we first considered
\begin{equation}
\int_{0}^{L}dy\frac{y}{\sqrt{y^{2}+d^{2}}}
\end{equation}
(We will take $L\rightarrow\infty$ momentarily.) Letting $u=y^{2}$,
\begin{align}
\int_{0}^{L}dy\frac{y}{\sqrt{y^{2}+d^{2}}} & =\frac{1}{2}\int_{0}^{L^{2}}du\frac{u}{\sqrt{u+d^{2}}}\\
 & =\left.\sqrt{u+d^{2}}\right|_{u=0}^{u=L^{2}}\\
 & =\sqrt{L^{2}+d^{2}}-d\\
 & =L\Big(1+\frac{1}{2}\big(\frac{d}{L}\big)^{2}+...\Big)-d
\end{align}
In the thermodynamic limit (\textit{i.e.}, large $L/d$),
\begin{equation}
\int_{0}^{L}dy\frac{y}{\sqrt{y^{2}+d^{2}}}=L-d
\end{equation}
so
\begin{equation}
\lim_{L\rightarrow\infty}\int_{0}^{L}dy\Big\{1-\frac{y}{\sqrt{y^{2}+d^{2}}}\Big\}=d
\end{equation}
We added the convergence factor without effect:
\begin{equation}
\lim_{\mu\rightarrow0}\int_{0}^{\infty}dy\Big\{\exp[-\mu y]-\frac{y\exp[-\mu y]}{\sqrt{y^{2}+d^{2}}}\Big\}=d
\end{equation}
Therefore,
\begin{equation}
\bra{\textbf{k}_{1}\xi_{1}\textbf{k}_{2}\xi_{2}}V\ket{\textbf{k}_{3}\xi_{3}\textbf{k}_{4}\xi_{4}} =\frac{2\pi d}{\epsilon A} \delta_{\xi_{1},\xi_{3}}\delta_{\xi_{2},\xi_{4}}\delta_{\textbf{k}_{1}+\textbf{k}_{2},\textbf{k}_{3}+\textbf{k}_{4}}
\end{equation}

For the $q\neq0$ case, we set $\mu=0$ in Eq.~(\ref{coulomb matrixele theta y}) and introduce
a Bessel function of order zero to get
\begin{align}
\bra{\textbf{k}_{1}\xi_{1}\textbf{k}_{2}\xi_{2}}V\ket{\textbf{k}_{3}\xi_{3}\textbf{k}_{4}\xi_{4}} & =\frac{2\pi }{\epsilon A} \delta_{\xi_{1},\xi_{3}}\delta_{\xi_{2},\xi_{4}}\delta_{\textbf{k}_{1}+\textbf{k}_{2},\textbf{k}_{3}+\textbf{k}_{4}}\int_{0}^{\infty}dyJ_{0}(qy) \nonumber \\
& \times \Big\{1-\frac{y}{\sqrt{y^{2}+d^{2}}}\Big\} \quad \text{for }q\neq 0 \\
 & =\frac{2\pi }{\epsilon A} \delta_{\xi_{1},\xi_{3}}\delta_{\xi_{2},\xi_{4}}\delta_{\textbf{k}_{1}+\textbf{k}_{2},\textbf{k}_{3}+\textbf{k}_{4}}\Big\{\frac{1}{q}-\frac{\exp[-qd]}{q}\Big\}
\end{align}
Putting everything together and reintroducing the factors of $e^2/(4\pi\epsilon_0\epsilon)$, our second-quantized Coulomb interaction was
\begin{align}
V & =\frac{e^2\pi }{4\pi\epsilon_0\epsilon A}\sum_{\textbf{k}\neq0}\sum_{\textbf{k},\textbf{p}}\sum_{\xi_{1},\xi_{2}}\Big\{\frac{1}{q}a_{\textbf{k}+\textbf{q},\xi_{1}}^{\dagger}a_{\textbf{p}-\textbf{q},\xi_{2}}^{\dagger}a_{\textbf{p},\xi_{2}}a_{\textbf{k},\xi_{1}}+\frac{1}{q}b_{\textbf{k},\xi_{1}}^{\dagger}b_{\textbf{p},\xi_{2}}^{\dagger}b_{\textbf{p}-\textbf{q},\xi_{2}}b_{\textbf{k}+\textbf{q},\xi_{1}}\\
 & -\frac{2\exp[-qd]}{q}a_{\textbf{k}+\textbf{q},\xi_{1}}^{\dagger}b_{\textbf{p},\xi_{2}}^{\dagger}b_{\textbf{p}-\textbf{q},\xi_{2}}a_{\textbf{k},\xi_{1}}\Big\}+\frac{2\pi e^2N^{2}d}{4\pi\epsilon_0\epsilon A}\nonumber 
\end{align}
where the last term is the capacitor energy, $a^\dagger_{\textbf{k},\xi_i}$ ($a_{\textbf{k},\xi_i}$) creates (destroys) an electron with wavevector $\textbf{k}$ and spin $\xi_i$, and $b^\dagger_{\textbf{k},\xi_i}$ ($b_{\textbf{k},\xi_i}$) creates (destroys) a hole with wavevector $\textbf{k}$ and spin $\xi_i$.

\subsubsection{Correlation energy}
For our general ($d\neq0$) two-component system, the correlation energy was given by \cite{InagakiKatayama}
\begin{equation}
\mathcal{E}_{\text{corr}}=\frac{\hbar iA}{2(2\pi)^3}\int_{0}^{1}\frac{d\lambda}{\lambda} \int d^{2}k\int d\omega\left[\begin{array}{cc}
\Pi_{e} & \Pi_{h}\end{array}\right]\left[\begin{array}{cc}
U_{\rm ee}^\lambda \lambda U_{\rm ee}^{0} & U_{\rm eh}^\lambda \lambda U_{\rm eh}^{0}\\
U_{\rm eh}^\lambda \lambda U_{\rm eh}^{0} & U_{\rm hh}^\lambda \lambda U_{\rm hh}^{0}
\end{array}\right]\left[\begin{array}{c}
\Pi_{e}\\
\Pi_{h}
\end{array}\right]\label{eq: ecorr def}
\end{equation}
where
\begin{equation}
U_{ij}^{0}=\begin{cases}
\frac{2\pi e^2}{k} & i=j\\
-\frac{2\pi e^2}{k}\exp[-kd] & i\neq j
\end{cases}
\end{equation}
and $U_{ij}^\lambda$ are the dynamically-screened effective interactions. 

$\Pi_i$ is the two-dimensional Lindhard polarizability for particle $i$. \cite{KuramotoKamimura} The real part is
\begin{equation}
\text{Re}\Pi_i(k,\omega)=\frac{m_i}{\hbar^{2}\pi}\Big[-1-\frac{\omega-k^{2}}{2k^{2}}\sqrt{1-\big(\frac{2k}{\omega-k^{2}}\big)^{2}}+\frac{\omega+k^{2}}{2k^{2}}\sqrt{1-\big(\frac{2k}{\omega+k^{2}}\big)^{2}}\Big]
\end{equation}
and the imaginary part is
\begin{equation}
\text{Im}\Pi_i(k,\omega)=-\frac{m_i}{\hbar^{2}\pi k}\Big[\sqrt{1-\big(\frac{\omega-k^{2}}{2k}\big)^{2}}-\sqrt{1-\big(\frac{\omega+k^{2}}{2k}\big)^{2}}\Big]
\end{equation}
In both expressions, each square root must be set to zero when it acquires a negative argument.  

Within the random-phase approximation,\cite{FetterWalecka} the effective interactions were given by the following Dyson-like or Ornstein-Zernike-like equations
\begin{equation}
U_{\rm ee}^\lambda=\lambda U_{\rm ee}^{0}+\lambda  U_{\rm ee}^{0}\Pi_{\rm e}U_{\rm ee}^\lambda+\lambda  U_{\rm eh}^{0}\Pi_{\rm h}U_{\rm eh}^\lambda 
\end{equation}
\begin{equation}
U_{\rm hh}^\lambda=\lambda  U_{\rm hh}^{0}+\lambda  U_{\rm eh}^{0}\Pi_{\rm e}U_{\rm eh}^\lambda +\lambda  U_{\rm hh}^{0}\Pi_{\rm h}U_{\rm hh}^\lambda 
\end{equation}
\begin{equation}
U_{\rm eh}^\lambda=\lambda U_{\rm eh}^{0}+\lambda U_{\rm ee}^{0}\Pi_{\rm e}U_{\rm eh}^\lambda +\lambda U_{\rm eh}^{0}\Pi_{\rm h}U_{\rm hh}^\lambda 
\end{equation}
or in matrix form,
\begin{equation}
\left[\begin{array}{cc}
U_{\rm ee}^\lambda  & U_{\rm eh}^\lambda \\
U_{\rm eh}^\lambda  & U_{\rm hh}^\lambda 
\end{array}\right]=\lambda  \left[\begin{array}{cc}
U_{\rm ee}^{0} & U_{\rm eh}^{0}\\
U_{\rm eh}^{0} & U_{\rm hh}^{0}
\end{array}\right]+\lambda \left[\begin{array}{cc}
\Pi_{\rm e}U_{\rm ee}^{0} & \Pi_{\rm h}U_{\rm eh}^{0}\\
\Pi_{\rm e}U_{\rm eh}^{0} & \Pi_{\rm h}U_{\rm hh}^{0}
\end{array}\right]\left[\begin{array}{cc}
U_{\rm ee}^\lambda  & U_{\rm eh}^\lambda \\
U_{\rm eh}^\lambda  & U_{\rm hh}^\lambda 
\end{array}\right].
\end{equation}

The correlation energy was rewritten as
\begin{align}
\mathcal{E}_{\text{corr}}=& \frac{\hbar iA}{2(2\pi)^3}\int_{0}^{1}\frac{d\lambda}{\lambda}U^0_{\rm ee} \int d^{2}k\int d\omega\left[\begin{array}{cc}
\Pi_{\rm e} & \Pi_{\rm h}\end{array}\right] \nonumber \\
 & \times \left[\begin{array}{cc}
U_{\rm ee} & -U_{\rm eh}\exp [-kd]\\
-U_{\rm eh}\exp [-kd] & U_{\rm hh}
\end{array}\right]\left[\begin{array}{c}
\Pi_{\rm e}\\
\Pi_{\rm h}
\end{array}\right]
\end{align}
The matrix of effective interactions was written in terms of the bare electron-electron interaction. Substitution of this result led to
\begin{align}
& \mathcal{E}_{\text{corr}}  =\frac{\hbar iA}{2(2\pi)^3}\int d^{2}k\int d\omega\int_{0}^{1}\frac{d\lambda}{\lambda}\frac{(\lambda U_{\rm ee}^{0})^2}{1-\lambda \big(\Pi_{\rm e}+\Pi_{\rm h}\big)U_{\rm ee}^{0}+\lambda ^2\big(1-\exp[-4kd]\big)\Pi_{\rm e}\Pi_{\rm h}\big(U_{\rm ee}^{0}\big)^{2}}\label{eq: final eq}\\
\times & \left[\begin{array}{cc}
\Pi_{\rm e} & \Pi_{\rm h}\end{array}\right] \left[\begin{array}{cc}
1+\lambda \big(\exp[-4kd]-1\big)\Pi_{\rm h}U_{\rm ee}^{0} & \exp[-2kd]\\
\exp[-2kd] & 1+\lambda \big(\exp[-4kd]-1\big)\Pi_{\rm e}U_{\rm ee}^{0}
\end{array}\right]\left[\begin{array}{c}
\Pi_{\rm e}\\
\Pi_{\rm h}
\end{array}\right]\nonumber 
\end{align}

To express this in $Ry_{\rm ex}$, we scaled lengths by $a_{\rm ex}$, wavevectors by $k_\text{F}$, and frequencies by $\hbar^2 k_{\text{F}}^{2}/(2m_{\rm red})$. Thus, the correlation energy per electron was
\begin{align}
& E_{\text{corr}}  =\frac{i}{\pi r_{s}^{2}}\int_{0}^{\infty}dk\int_{0}^{\infty}d\omega\int_{0}^{1}d\lambda  \frac{k\lambda \big(U_{ee}^{0}\big)^2}{1-\lambda\big(\Pi_{e}+\Pi_{h}\big)U_{ee}^{0}+\lambda^{2}\big(1-\exp[-4\sqrt{2}kd/r_{s}]\big)\Pi_{e}\Pi_{h}\big(U_{ee}^{0}\big)^{2}}\nonumber\\
\times & \left[\begin{array}{cc}
\Pi_{e} & \Pi_{h}\end{array}\right] \left[\begin{array}{cc}
1+\lambda\big(\exp[-4\sqrt{2}kd/r_{s}]-1\big)\Pi_{h}U_{ee}^{0} & \exp[-2\sqrt{2}kd/r_{s}]\\
\exp[-2\sqrt{2}kd/r_{s}] & 1+\lambda\big(\exp[-4\sqrt{2}kd/r_{s}]-1\big)\Pi_{e}U_{ee}^{0}
\end{array}\right] \nonumber\\
&\times\left[\begin{array}{c}
\Pi_{e}\\
\Pi_{h}
\end{array}\right]Ry_{\rm ex}  
\end{align}
The result of the matrix multiplication is
\begin{align}
 E_{\text{corr}} = & \frac{i}{\pi r_{s}^{2}}\int_{0}^{\infty}dk\int_{0}^{\infty}d\omega\int_{0}^{1}d\lambda k \frac{\lambda}{1-\lambda\big(\Pi_{\rm e}+\Pi_{\rm h}\big)U_{\rm ee}^{0}+\lambda^{2}\big(1-\exp[-4\sqrt{2}kd/r_{s}]\big)\Pi_{\rm e}\Pi_{\rm h}\big(U_{\rm ee}^{0}\big)^{2}}  \nonumber \\
 & \times \Big( U_{\rm ee}^{0}\big)^{2}\big(\Pi_{\rm e}^{2}+\Pi_{\rm h}^{2}\big)+2\big(U_{\rm ee}^{0}\big)^{2}\Pi_{\rm e}\Pi_{\rm h}\exp[-2\sqrt{2}kd/r_{s}] \nonumber \\
 & +\lambda\big(\exp[-4\sqrt{2}kd/r_{s}]-1\big)\big(U_{\rm ee}^{0}\big)^{3}\big(\Pi_{\rm e}+\Pi_{\rm h}\big)\Pi_{\rm e}\Pi_{\rm h} \Big) Ry_{\rm ex}  \label{final dimensionless}
\end{align}

\subsection{Chemical potentials}
The chemical potentials for excitons and free carriers included ideal and capacitor contributions, both of which were evaluated exactly. However, for electrons and holes, there were additional exchange and correlation contributions that were evaluated numerically. In this case, we first evaluated Helmholtz free energies as a function of density, fit these functions to a simple form, and then computed derivatives to obtain the exchange-correlation chemical potentials. 

\subsubsection{Ideal chemical potentials}
The chemical potential for ideal fermions or bosons in two dimensions can be derived starting from the partition function
\begin{equation}
\Xi=\sum_{n_{1},n_{2},...,n_{k},...}\exp\Big[-\beta\sum_{k}(\varepsilon_{k}-\mu)n_{k}\Big]
\label{general partition fn}
\end{equation}
where $n_{k}$ are occupation numbers and $\varepsilon_{i,k}$ are the single particle energies given by
\begin{equation}
    \varepsilon_{i,k}=\frac{\hbar^2 k^2}{2m_i}.
    \label{single particle energies}
\end{equation}
For fermions, factorizing the exponential in Eq.~(\ref{general partition fn}) gives
\begin{align}
\Xi & =\prod_{k}\Big\{\sum_{n_{k}=0}^{1}\exp\Big[-\beta(\epsilon_{k}-\mu)n_{k}\Big]\Big\}\label{eq: part fn fermion}\\
 & =\prod_{k}\Big\{1+\exp\Big[-\beta(\epsilon_{k}-\mu)\Big]\Big\}
\end{align}
The total number of particles was given by
\begin{align}
\langle N\rangle & =\frac{\partial\ln\Xi}{\partial(\beta\mu)}\\
 & =\sum_{k}\frac{\exp[\beta(\mu-\varepsilon_k)]}{1+\exp[\beta(\mu-\varepsilon_k)]}\\
 & =\sum_{k}f(k,\beta,\mu)\label{eq: total n fermion}
\end{align}
where $f$ is the Fermi-Dirac distribution. Moving to the thermodynamic limit, we replaced the sum by an integral and wrote
\begin{align}
\langle N\rangle & =\frac{\xi_iA}{(2\pi)^{2}}\int d^{2}k\frac{\exp[\beta(\mu-\varepsilon_{i,k})]}{1+\exp[\beta(\mu-\varepsilon_{i,k})]}\\
 & =\frac{\xi_iA}{2\pi}\int_{0}^{\infty}\frac{\exp[\beta(\mu-\varepsilon_{i,k})]}{1+\exp[\beta(\mu-\varepsilon_{i,k})]}kdk
\end{align}
where $\xi_i=2$ is the spin degeneracy for particle $i$. We evaluated the integral using a $u$-substitution: $u=\exp[\beta(\mu-\varepsilon_{i,k})]$.
\begin{align}
\langle N_i\rangle & =\frac{\xi_iA}{2\pi}\frac{m_i}{\hbar^{2}\beta}\int_{0}^{\exp[\beta\mu]}\frac{1}{1+y}dy\\
 & =\frac{\xi_iA}{\lambda_{i}^{2}}\ln\big(1+\exp[\beta\mu]\big)
\end{align}
where we have the thermal de Broglie wavelength
\begin{equation}
\lambda_{i}^{2}=\frac{2\pi\hbar^{2}\beta}{m_i} \label{thermal wavelength}
\end{equation}
After some algebra, we obtained the chemical potential for non-interacting fermions in two dimensions:
\begin{equation}
\beta\mu_0^i=\ln\big(\exp[n\lambda_{i}^{2}/(\xi_i)]-1\big).
\label{ideal chempot fermions}
\end{equation}

For bosons, the grand partition function was written as
\begin{equation}
\Xi=\prod_{k}\Big(1-\exp[\beta(\mu-\varepsilon_k)]\Big)^{-1}
\end{equation}
Following the same steps as above, we get
\begin{align}
\langle N_i\rangle & =\xi_i\sum_{k}\frac{\exp[\beta(\mu-\varepsilon_{i,k})]}{1-\exp[\beta(\mu-\varepsilon_{i,k})]}\\
 & =\frac{\xi_iA}{2\pi}\frac{m_i}{\hbar^{2}\beta}\int_{0}^{\exp[\beta\mu]}\frac{1}{1-y}dy\\
 & =-\frac{\xi_iA}{\lambda_{i}^{2}}\ln\big(1-\exp[\beta\mu]\big)
\end{align}
Hence, we obtained the chemical potential for bosons:
\begin{equation}
\beta\mu_0^i=\ln\big(1-\exp[-n_i\lambda_{i}^{2}/\xi_i]\big).
\label{ideal chempot bosons}
\end{equation}

\subsubsection{Exchange free energy}
The first-order exchange contribution to the thermodynamic potential was given by\cite{FetterWalecka,Phatisena} 
\begin{equation}
\Omega_{\rm exch}(\beta,\mu)=-\sum_{i=\text{e,h}}\frac{\xi_i}{2A}\sum_{\textbf{k},\textbf{p}}\frac{2\pi e^{2}}{4\pi\epsilon_0\epsilon|\textbf{k}-\textbf{p}|}f_{i}(k,\beta,\mu)f_{i}({p},\beta,\mu)\label{eq: exchange thermo pot}
\end{equation}
where $f_i$ is the Fermi-Dirac distribution for particle $i$ and $\beta^{-1}=k_{\rm B}T$. This equation is exact when we use the chemical potential for the interacting system, $\mu$. Following previous work,\cite{RustagiKemper,Phatisena} we replaced this exact chemical potential by the one corresponding to the non-interacting system, Eq.~(\ref{ideal chempot fermions}). Thus, our Helmholtz exchange free energy was
\begin{equation}
\mathcal{F}_{\rm exch}(\beta,n)=-\sum_{i=e,h}\frac{\xi_i}{2A}\sum_{\textbf{k},\textbf{p}}\frac{2\pi e^{2}}{4\pi\epsilon_0\epsilon|\textbf{k}-\textbf{p}|}f_{i}\big(k,\beta,\mu_0(n)\big)f_{i}\big({p},\beta,\mu_0(n)\big)
\end{equation}

Moving to dimensionless quantities, we let $k'=k/k_\text{F}$ and $t=T/T_\text{F}$, where the Fermi temperature is
\begin{align}
T_{\text{F}} & =\frac{\hbar^{2}k_{\text{F}}^{2}}{2m_{\rm red}k_\text{B}}
\end{align}
In these quantities, the ideal chemical potentials became
\begin{equation}
\beta\mu_{0}^{i}=\begin{cases}
\ln\Big[\exp\big(\frac{1}{t(\sigma+1)}\big)-1\Big] & i=\text{e}\\
\ln\Big[\exp\big(\frac{\sigma}{t(\sigma+1)}\big)-1\Big] & i=\text{h}
\end{cases}
\end{equation}
and the Fermi-Dirac distributions become
\begin{equation}
f_{i}(k',t,n)=\begin{cases}
\frac{\exp\big[1/\big(t(\sigma+1)\big)\big]-1}{\exp\big[k'^{2}/\big(t(\sigma+1)\big)\big]+\exp\big[1/\big(t(\sigma+1)\big)\big]-1} & i=e\\
\frac{\exp\big[\sigma/\big(t(\sigma+1)\big)\big]-1}{\exp\big[k'^{2}\sigma/\big(t(\sigma+1)\big)\big]+\exp\big[\sigma/\big(t(\sigma+1)\big)\big]-1} & i=h
\end{cases}\label{eq: reduced fermi fn}
\end{equation}
Note that when the electrons and holes have different masses, they experience two different effective temperatures: 
\begin{equation}
t_i=\begin{cases}
t(\sigma+1) & i=\text{e}\\
t(\sigma+1)/\sigma & i=\text{h}
\end{cases}\label{eq: effective temp}
\end{equation}
In these reduced units, the density dependence entered only through the reduced temperature. 

Converting the sums to integrals, 
\begin{equation}
\frac{\mathcal{F}_{\rm exch}(t)}{N}=-\sum_{i=\text{e,h}}\frac{k_{\rm F}e^{2}}{(2\pi)^{2}}\int d^{2}k'\int d^{2}p'\frac{f_{i}(k',t)f_{i}(p',t)}{4\pi\epsilon_0\epsilon|\textbf{k}'-\textbf{p}'|}
\end{equation}
Phatisena and co-workers \cite{Phatisena} expressed the exchange free energy of a two-dimensional electron gas relative to its exchange energy at zero temperature, $E_{\rm exch}$:
\begin{align}
    \frac{\mathcal{F}_{\rm exch}^{2DEG}(t)}{\mathcal{E}_{\rm exch}} & =3\int_0^\infty dx x^2 f(x,t) \int_0^1 dz f(xz,t)K(z) \label{2deg exch}\\
    & = C(t)
\end{align}
where $K(z)$ is the complete elliptic integral of the first kind. To apply this result for our two-component system, we evaluated it at the previously defined effective temperatures:
\begin{equation}
    F_{\rm exch}=-\sum_{i=\text{e,h}}\frac{1.2004}{r_s}C(t_i)Ry_{\rm ex}
\end{equation}

\subsubsection{Correlation free energy}
Using the linked cluster theorem,\cite{Mahan} the correlation thermodynamic potential for a bilayer electron-hole liquid obtained by summing ring diagrams\cite{FetterWalecka} was expressed as
\begin{align}
&\Omega_{\rm corr}(\beta,\mu) =-\frac{1}{2\beta}\sum_{\omega_{l},k}\int_{0}^{1}\frac{d\lambda}{\lambda}\left[\begin{array}{cc}
\Pi_{\rm e} & \Pi_{\rm h}\end{array}\right] \left[\begin{array}{cc}
U_{\rm ee}(\lambda,k)U_{\rm ee}^{0}(\lambda,k) & U_{\rm eh}(\lambda,k)U_{\rm eh}^{0}(\lambda,k)\\
U_{\rm eh}(\lambda,k)U_{\rm eh}^{0}(\lambda,k) & U_{\rm hh}(\lambda,k)U_{\rm hh}^{0}(\lambda,k)
\end{array}\right]\nonumber\\
&\times\left[\begin{array}{c}
\Pi_{\rm e}(k,\omega_{l},\beta)\\
\Pi_{\rm h}(k,\omega_{l},\beta)
\end{array}\right]\\
 & =-\frac{A}{2\beta(2\pi)^{2}}\sum_{\omega_{l}}\int d^{2}k\int_{0}^{1}d\lambda U_{\rm ee}^{0}\frac{\lambda U_{\rm ee}^{0}}{1-\lambda\big(\Pi_{\rm e}+\Pi_{\rm h}\big)U_{\rm ee}^{0}+\lambda^{2}\big(1-\exp[-4kd]\big)\Pi_{\rm e}\Pi_{\rm h}\big(U_{\rm ee}^{0}\big)^{2}}\nonumber \\
 & \times\left[\begin{array}{cc}
\Pi_{\rm e} & \Pi_{\rm h}\end{array}\right]\left[\begin{array}{cc}
1+\lambda\big(\exp[-4kd]-1\big)\Pi_{\rm h}U_{\rm ee}^{0} & \exp[-2kd]\\
\exp[-2kd] & 1+\lambda\big(\exp[-4kd]-1\big)\Pi_{\rm e}U_{\rm ee}^{0}
\end{array}\right]\left[\begin{array}{c}
\Pi_{\rm e}\\
\Pi_{\rm h}
\end{array}\right]
\end{align}
In lieu of integrating over a continuous frequency, we now summed over discrete bosonic Matsubara frequencies
\begin{equation}
    \omega_l=\frac{2\pi l}{\beta} \quad l=0,\pm1,\pm2,...
\end{equation}
Additionally, the static ($l=0$) and dynamic ($l\neq0$) finite-temperature polarizabilities were \cite{Phatisena}
\begin{equation}
\Pi_j(k,l=0,t)=-\frac{4m_j}{\pi\hbar^{2}k}\int_{0}^{k/2}dx\frac{xf(x,t_j)}{\sqrt{k^{2}-4x^{2}}}
\end{equation}
and
\begin{equation}
\Pi_j(k,l\neq0,t)=-\frac{4m_j}{\pi\hbar^{2}}\int_{0}^{\infty}dx\frac{xf_j(x,t)\cos\phi}{\Big[\big(k^{4}-4x^{2}k^{2}-4l^{2}\pi^{2}t_j^{2}\big)^{2}+16l^{2}\pi^{2}t_j^{2}k{}^{4}\Big]^{1/4}}
\end{equation}
where
\begin{equation}
\tan(2\phi)=\frac{4k^{2}l\pi t_j}{k^{4}-4x^{2}k^{2}-4l^{2}\pi^{2}t_j^{2}}
\end{equation}
When computing $\phi$, we used ``atan2($y,x$),'' the two-argument arctangent function which yields the angle between the positive $x$-axis and the point ($x,y$). 
Using the same dimensionless quantities as before,
\begin{align}
& F_{\rm corr}(t,r_{s}) =-\frac{t}{r_{s}^{2}}\sum_{\omega_{l}}\int_{0}^{\infty}dk\int_{0}^{1}d\lambda \nonumber\\
&\times\Big\{ \frac{k\lambda}{1-\lambda\big(\Pi_{\rm e}+\Pi_{\rm h}\big)U_{\rm ee}^{0}+\lambda^{2}\big(1-\exp[-4\sqrt{2}kd'/r_{s}]\big)\Pi_{\rm e}\Pi_{\rm h}\big(U_{\rm ee}^{0}\big)^{2}} \nonumber\\
& \times \Big(\big(U_{\rm ee}^{0}\big)^{2}\big(\Pi_{\rm e}^{2}+\Pi_{\rm h}^{2}\big)+2\big(U_{\rm ee}^{0}\big)^{2}\Pi_{\rm e}\Pi_{\rm h}\exp[-2\sqrt{2}kd'/r_{s}]\nonumber\\
& +\lambda\big(\exp[-4\sqrt{2}kd'/r_{s}]-1\big)\big(U_{\rm ee}^{0}\big)^{3}\big(\Pi_{\rm e}+\Pi_{\rm h}\big)\Pi_{\rm e}\Pi_{\rm h}\Big) \Big) \Big\} Ry_{\rm ex} 
\end{align}

\subsubsection{Fitting exchange-correlation free energies}
To obtain smooth exchange-correlation chemical potentials, we fit the exchange-correlation free energies $F_{\rm XC}=F_{\rm exch}+F_{\rm corr}$ to a function which has a smooth and simple derivative. Our target function was
\begin{equation}
F_{\rm fit}(n;n_{\rm C})=\begin{cases}
\sum_{j=0}^{N}a_{j}n^{j/2} & n<n_{\rm C}\\
\sum_{j=N+1}^{2N+2}a_{j}(\ln n)^{j-N-1} & n\geq n_{\rm C}
\end{cases} \label{fit fn}
\end{equation}
$N$ is the order of both polynomials and $n_{\rm C}$ is the density cutoff (typically 10$^{9}$ cm$^{-2}$) above which we fit the data using a polynomial of $\ln n$.
By defining
\begin{equation}
c_{j}(n;n_{C})=\begin{cases}
n^{j/2} & j\leq N\text{ and }n<n_{C}\\
0 & j\leq N\text{ and }n\geq n_{C}\\
0 & j>N\text{ and }n<n_{C}\\
(\ln n)^{j-N-1} & j>N\text{ and }n\geq n_{C}
\end{cases}
\end{equation}
and $\vec{a}$ as our vector of desired fitting coefficients, Eq.~\ref{fit fn} was written as
\begin{equation}
F_{\rm fit}(n;n_{\rm C})=\vec{c}(n;n_{\rm C})\cdot\vec{a}
\end{equation}

At $n_{\rm C}$, we required the values and first derivatives of the polynomials to the left and right to be equal:
\begin{equation}
\sum_{j=0}^{N}a_{j}(n_{C})^{j/2}=\sum_{j=N+1}^{2N+2}a_{j}(\ln n_{C})^{j-N-1}
\end{equation}
\begin{equation}
\frac{d}{dn}\left(\sum_{j=0}^{N}a_{j}(n)^{j/2}\right)_{n=n_{C}}=\frac{d}{dn}\left(\sum_{j=N+1}^{2N+2}a_{j}(\ln n)^{j-N-1}\right)_{n=n_{C}}
\end{equation}
By defining
\begin{equation}
d_{1j}=\begin{cases}
n_{C}^{j/2} & j\leq N\\
-(\ln n_{C})^{j-N-1} & j>N
\end{cases}
\end{equation}
\begin{equation}
d_{2j}=\begin{cases}
\frac{j}{2}n_{\rm C}^{j/2-1} & j\leq N\\
-\frac{(j-N-1)}{n_{\rm C}}(\ln n_{\rm C})^{j-N-2} & j>N
\end{cases}
\end{equation}
our constraints became
\begin{equation}
\vec{d}_{1}\cdot\vec{a}=0
\end{equation}
\begin{equation}
\vec{d}_{2}\cdot\vec{a}=0
\end{equation}

To minimize the squared residual error
\begin{equation}
\mathcal{E}=\sum_{\alpha=1}^{N_{\text{density}}}\big(F_{\rm fit}(n_{\alpha};n_{C})-F_{\rm XC}(n_{\alpha})\big)^{2}
\end{equation}
subject to our constraints, we introduced two Lagrange multipliers, $\lambda_{1}$ and $\lambda_{2}$. Our Lagrangian was
\begin{equation}
\mathcal{L}(\{n_{\alpha}\},n_{\rm C},\lambda_{1},\lambda_{2})=\sum_{\alpha=1}^{N_{\text{density}}}\big(F_{\rm fit}(n_{\alpha};n_{\rm C})-F_{\rm XC}(n_{\alpha})\big)^{2}-\lambda_{1}\vec{d_{1}}\cdot\vec{a}-\lambda_{2}\vec{d_{2}}\cdot\vec{a}
\end{equation}
To find $\vec{a}$ which minimizes $\mathcal{L}$, we set its derivative equal to 0:
\begin{align}
\frac{\partial\mathcal{L}}{\partial a_{i}} & =0\\
 & =2\sum_{\alpha=1}^{N_{\text{density}}}\big(F_{fit}(n_{\alpha};n_{C})-F_{XC}(n_{\alpha})\big)^{2}\big)\left.\frac{\partial F_{fit}}{\partial a_{i}}\right|_{n=n_{\alpha}}-\lambda_{1}d_{1i}-\lambda_{2}d_{2i}\label{eq: dL/da}
\end{align}
Note that
\begin{equation}
\left.\frac{\partial F_{\rm fit}}{\partial a_{i}}\right|_{n=n_{\alpha}}=c_{i}(n_{\alpha};n_{\rm C})
\end{equation}
Defining
\begin{equation}
\chi_{ij}^{-1}=2\sum_{\alpha=1}^{N_{\text{density}}}c_{i}(n_{\alpha};n_{\rm C})c_{j}(n_{\alpha};n_{\rm C})
\end{equation}
and
\begin{equation}
b_{i}=2\sum_{\alpha=1}^{N_{\text{density}}}F_{\rm XC}(n_{\alpha})c_{i}(n_{\alpha})
\end{equation}
(\ref{eq: dL/da}) became
\begin{equation}
\chi^{-1}\cdot\vec{a}=\vec{b}+\lambda_{1}\vec{d_{1}}+\lambda_{2}\vec{d_{2}}
\end{equation}
Our desired coefficients were given by
\begin{equation}
\vec{a}=\chi\cdot\vec{b}+\lambda_{1}\chi\cdot\vec{d_{1}}+\lambda_{2}\chi\cdot\vec{d_{2}}
\end{equation}

The Lagrange multipliers were determined by
\begin{align}
0 & =\vec{d_{1}}\cdot\vec{a}\\
 & =\vec{d_{1}}\cdot\chi\cdot\vec{b}+\lambda_{1}\vec{d}_{1}\cdot\chi\cdot\vec{d_{1}}+\lambda_{2}\vec{d}_{1}\cdot\chi\cdot\vec{d_{2}}
\end{align}
\begin{align}
0 & =\vec{d_{2}}\cdot\vec{a}\\
 & =\vec{d_{2}}\cdot\chi\cdot\vec{b}+\lambda_{1}\vec{d}_{2}\cdot\chi\cdot\vec{d_{1}}+\lambda_{2}\vec{d}_{2}\cdot\chi\cdot\vec{d_{2}}
\end{align}
By defining
\begin{equation}
g_{ij}=\vec{d}_{i}\cdot\chi\cdot\vec{d}_{j}
\end{equation}
and
\begin{equation}
h_{i}=-\vec{d}_{i}\cdot\chi\cdot\vec{b}
\end{equation}
we have
\begin{equation}
G\cdot\vec{\lambda}=\vec{h}
\end{equation}
so
\begin{equation}
\vec{\lambda}=G^{-1}\cdot\vec{h}
\end{equation}

\begin{figure}
    \centering
    \includegraphics[width=12cm]{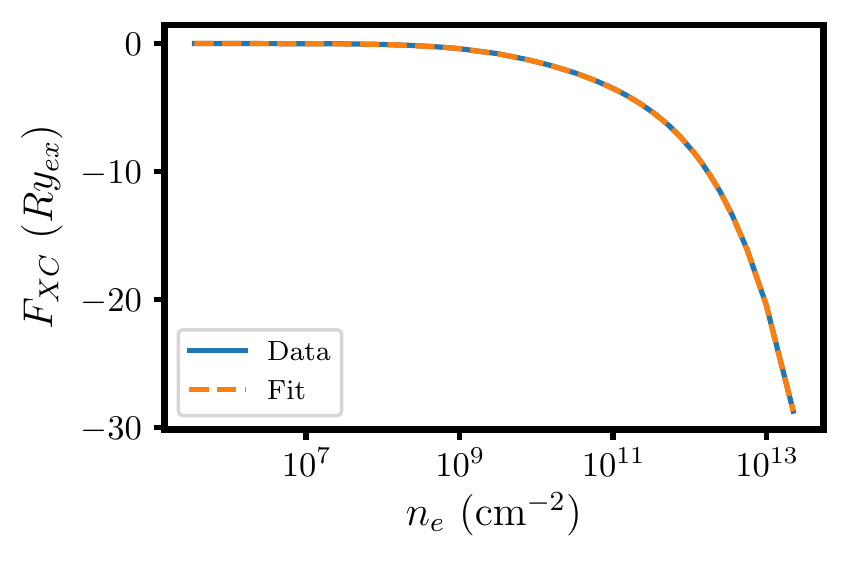}
    \caption{Exchange and correlation contributions to the Helmholtz free energy of a bilayer EHL with $\sigma=0.1$, $d=1.5a_{\rm ex}$, and $T=8\textrm{K}$. The x-axis is the density of free carriers in units of number of electrons per cm$^{2}$.}
    \label{fig:s2}
\end{figure}

Figure~\ref{fig:s2} shows the results of the fit compared to the raw data for a bilayer EHL with $\sigma=0.1$, $d=1.5a_{\rm ex}$, and $T=8\textrm{K}$. The deviation is at most 0.02 $Ry_{\rm ex}$.

\begin{figure}[!h]
\centering
   \includegraphics[width=12cm]{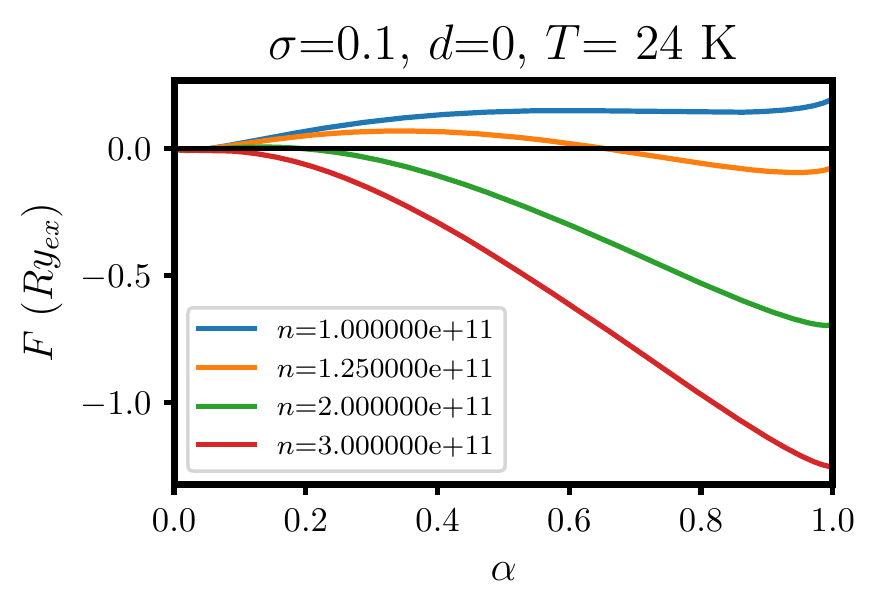}
    \caption{Helmholtz free energy as a function of ionization ratio. The total excitation densities in the legend are in units of number of electrons per cm$^{2}$. The Mott density is $n_{\rm Mott} \lesssim 1.25\times10^{11}$ cm$^{-2}$.}
    \label{fig:s3}
\end{figure}

\subsection{Solving the law of mass action}
To evaluate the ionization ratio $\alpha$ at low densities, we solved the law of mass action given by Eqs. 11 and 12 in the main text using a root-finding algorithm: the bisection method. However, multiple roots appeared at higher densities (shown below). The correct choice was the one that minimizes the system's free energy, so we first computed the change in Helmholtz free energy due to carriers partitioning into free and bound states:
\begin{align}
dF & =\mu_{\rm eh}dn_{\rm eh}+\mu_{\rm X}dn_{\rm X}\\
 & =(\mu_{\rm eh}-\mu_{\rm X})n_{\rm tot}d\alpha.
\end{align}
Thus,
\begin{equation}
\frac{dF}{d\alpha}=(\mu_{\rm eh}-\mu_{\rm X})n_{\rm tot}
\end{equation}
Integrating from some arbitrary reference value $\alpha_{0}$ leads to
\begin{equation}
F(\alpha)=F(\alpha_{0})+n_{tot}\int_{\alpha_{0}}^{\alpha}\big(\mu_{tot}^{free}(\tilde{\alpha})-\mu_{tot}^{X}(\tilde{\alpha})\big)d\tilde{\alpha}
\end{equation}
Because the location of the minimum doesn't depend on $F(\alpha_{0})$ or the factor of $n_{tot}$, these terms were neglected. 

Fig.~\ref{fig:s3} shows the Helmholtz free energy as a function of the ionization ratio $\alpha$. When the total excitation density is 1.25 $\times10^{11}$ cm$^{-2}$, there are two extrema located at $\alpha \approx$0.35 and 0.95. While these are both solutions to the law of mass action, $\alpha\approx0.95$ is the correct choice as it minimizes the free energy.

\begin{figure}[!h]
\centering
    \includegraphics[width=12cm]{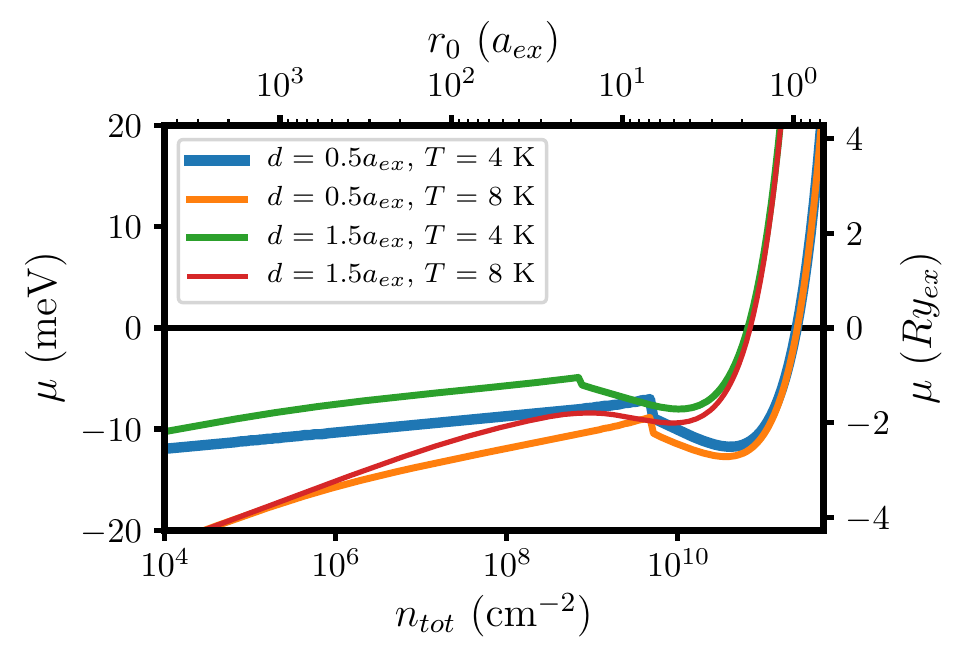}
    \caption{Chemical potentials for the electron-hole-exciton system with $\sigma=0.1$.}
    \label{fig:s4}
\end{figure}

\subsection{Maxwell equal-area constructions}
After calculating the ionization ratio $\alpha$ as a function of $n_{\rm tot}$, we obtained chemical potentials for the electron-hole-exciton system shown in Fig.~\ref{fig:s4}. These potentials violate thermodynamic stability for densities near $10^{10}$ cm$^{-2}$ since their first derivatives with respect to density are negative. We determined the liquid-gas coexistence densities using a standard Maxwell equal-area construction;\cite{Chandler} we identified the value of the chemical potential $\mu_{\rm constant}$ that encloses two regions of equal area.